\documentclass[useAMS, subeqn, usenatbib]{mn2e}
\usepackage{natbib}
\usepackage{graphicx}
\usepackage{epsfig}
\usepackage{hyperref}
\usepackage{amsmath}
\usepackage{amssymb}
\usepackage{color}
\usepackage{datetime}
\usepackage{textgreek}
\usepackage{pdflscape}
\usepackage{flafter}

\newdateformat{monthyeardate}{%
\monthname[\THEMONTH], \THEYEAR}

\title[VVV Astrometric Catalogue]{VIRAC: The VVV Infrared Astrometric Catalogue}
\date{\monthyeardate\today}

\author[L.C. Smith et al.]{L.C. Smith$^1$\thanks{\href{mailt:l.c.smith@herts.ac.uk}{l.c.smith@herts.ac.uk}}, P.W. Lucas$^1$, R. Kurtev$^{2,3}$, R. Smart$^{1,4}$, D. Minniti$^{3,5,6}$,
\newauthor
J. Borissova$^{2,3}$, H.R.A Jones$^1$, Z.H. Zhang$^7$, F. Marocco$^1$, C. Contreras Pe\~{n}a$^8$, 
\newauthor
M. Gromadzki$^9$, M.A. Kuhn$^{2,3}$, J.E. Drew$^1$, D.J. Pinfield$^1$, L.R. Bedin$^{10}$ \\
$^1$ Centre for Astrophysics Research, School of Physics, Astronomy and Mathematics, University of Hertfordshire, College Lane, \\Hatfield AL10 9AB, UK\\
$^2$ Instituto de F\'isica y Astronom\'ia, Universidad de Valpara\'iso, Av. Gran Breta\~na 1111, Playa Ancha, Casilla 5030, Valpara\'iso, Chile\\
$^3$ The Millennium Institute of Astrophysics (MAS), Av. Vicu\~na Mackenna 4860, 782-0436 Macul,  Santiago, Chile \\
$^4$ Istituto Nazionale di Astrofisica, Osservatorio Astronomico di Torino, Strada Osservatorio 20, 10025 Pino Torinese, Italy\\
$^5$ Departamento de Ciencias Fisicas, Universidad Andres Bello, Republica 220, Santiago, Chile \\
$^6$ Vatican Observatory, V-00120 Vatican City State, Italy\\
$^7$ GEPI, Observatoire de Paris, PSL Research University, CNRS, Universit{\'e} Paris Diderot, Sorbonne Paris Cit{\'e}, Place Jules Janssen, \\ 92195 Meudon, France \\
$^8$ Department of Physics and Astronomy, University of Exeter, Stocker Road, Exeter, Devon EX4 4SB, UK\\
$^9$ Warsaw University Astronomical Observatory, Al. Ujazdowskie 4, PL-00-478, Warszawa, Poland\\
$^{10}$ INAF-Osservatorio Astronomico di Padova, Vicolo dell'Osservatorio 5, I-35122 Padova, Italy\\
}

\begin{document}

\maketitle

\begin{abstract}
We present VIRAC version 1, a near-infrared proper motion and parallax catalogue of the VISTA VVV survey for 312,587,642 unique sources averaged across all overlapping pawprint and tile images covering 560 deg$^2$ of the bulge of the Milky Way and southern disk. The catalogue includes 119 million high quality proper motion measurements, of which 47 million have statistical uncertainties below 1~mas~yr$^{-1}$. In the $11<K_s<14$ magnitude range the high quality motions have a median uncertainty of $0.67$~mas~yr$^{-1}$. The catalogue also includes 6,935 sources with quality-controlled 5~$\sigma$ parallaxes with a median uncertainty of 1.1~mas. The parallaxes show reasonable agreement with the TYCHO-Gaia Astrometric Solution (TGAS), though caution is advised for data with modest significance. The SQL database housing the data is made available via the web. We give example applications for studies of Galactic structure, nearby objects (low mass stars and brown dwarfs, subdwarfs, white dwarfs) and kinematic distance measurements of YSOs. Nearby objects discovered include LTT 7251 B, an L7 benchmark companion to a G dwarf with over 20 published elemental abundances, a bright L sub-dwarf, VVV 1256-6202, with extremely blue colours and nine new members of the 25~pc sample. We also demonstrate why this catalogue remains useful in the era of Gaia. Future versions will be based on profile fitting photometry, use the Gaia absolute reference frame and incorporate the longer time baseline of the VVV extended survey (VVVX).
\end{abstract}


\section{Introduction}

In recent years the astronomical community has undertaken several large projects that aim to measure the structure and dynamics of the Milky Way, e.g. the Gaia astrometric mission \citep{gaia-cite}, the optical spectographs 4MOST \citep{4most-cite} and WEAVE \citep{weave-cite} and the infrared spectrographs APOGEE I-II (\citealt{apogee-cite-1}; \citealt{apogee-cite-2}) and MOONS \citep{moons-cite}. The VISTA Variables in the Via Lactea (VVV) survey \citep{minniti10} complements this effort by providing time series $K_s$ photometry of a 560 deg$^2$ region of the Galactic disc and bulge, much of which is hidden from the view of optical surveys. Although originally planned to measure 3D Galactic structure using standard candles (RR Lyrae, red clump giants, Cepheids) it has become apparent that VISTA has excellent astrometric properties that enable proper motion measurements across the Galaxy, either using VISTA alone (\citealt{libralato15}; \citealt{cioni16}), or in combination with other datasets \citep{cioni14} such as the Two Micron All Sky Survey (2MASS, \citealt{skrutskie06}).

VVV data comprise typically between 50 and 80 epochs of $K_s$ photometry over five years (2010 to 2015). In addition, VVV includes two epochs of Z, Y, J and H photometry taken at the beginning and end of the survey. Each epoch is sub-divided into independent images that are treated separately in our astrometry.

In this paper we present version 1 (V1) of the VVV Infrared Astrometric Catalogue (VIRAC), based on the standard products provided by the v1.3 pipeline of the Cambridge Astronomical Survey Unit (CASU). This VIRAC V1 catalogue provides \textit{relative} proper motions and parallaxes for all stars for which they could be measured in the individual pointings (pawprints) of VVV. The catalogue is available at \href{http://vvv.herts.ac.uk}{vvv.herts.ac.uk} and it will be made available in the VISTA Science Archive (VSA, \citealt{cross12}, see \href{http://horus.roe.ac.uk/vsa/}{horus.roe.ac.uk/vsa}) and the ESO Archive.

Following the release of the Gaia 2nd Data Release, the VVV team also plans to provide VIRAC V2, a deeper catalogue based on profile fitting photometry of the VVV dataset that will provide astrometry on the Gaia absolute reference frame. The work of \citet{contrerasramos17} is an excellent example of the power of profile fitting for NGC 6544. We will also explore the possibility of further increasing the depth of VIRAC using a shift-and-stack algorithm, as has been done by the ALLWISE project \citep{kirkpatrick14}. The VVV project has been extended by a new survey, VVVX, that continues to survey the original VVV area approximately 9 times up to 2020, while extending the area to cover an additional 1100 deg$^2$ of the Galactic disc and bulge at 25-40 epochs. We plan to incorporate VVVX data into future astrometric products.

Here we describe our proper motion and parallax methodology and present initial results. These include nearby high proper motion stars and brown dwarfs and an illustration of how VIRAC can be used at large distances across the Milky Way. Our results include a complete catalogue of visually confirmed sources with proper motion, $\mu > 200$~mas/yr, complementing the VVV high proper motion catalogue recently published by \citet{kurtev17} for relatively bright stars with magnitudes $K_s <13.5$.

In section \ref{data_description} we describe the dataset and data selection. Section \ref{astrometry} details the source matching and the proper motion and parallax calculations. In Section \ref{pmresults} we describe our quality checks and quality flags for the proper motion catalogue, using internal self-consistency, visual inspection and comparison to both Tycho-Gaia Astrometric Solution (TGAS, \citealt{tgas-cite}) and the results of \citet{kurtev17}. This section also describes our table of visually confirmed high proper motion stars (the table itself can be found in the appendices) and new proper motion companions to TGAS stars. In Section \ref{plxcat} we describe the parallax catalogue and parallax quality checks using TGAS. In section \ref{discoveries} we describe discoveries of note from the parallax dataset and high proper motion sources. In Section \ref{pan-galactic} we demonstrate applications at large Galactic distances, including measurement of the Galactic rotation curve and the motion of the Sagaittarius dwarf galaxy.

\section{Data description and selection}\label{data_description}

The VISTA Infrared Camera (VIRCAM) is the current largest near-infrared imager in astronomical use, consisting of sixteen 2048$\times$2048 pixel mercury cadmium telluride arrays. The VISTA/VIRCAM combination gives a total viewing area of 0.6 deg$^{2}$ for each pointing or "pawprint" of the telescope. Detectors are placed in a 4$\times$4 grid with spacing of 0.9 detector widths in the $y$ direction and 0.425 detector widths in the $x$ direction. The conventional tiling pattern used in VVV consists of six separate pawprints (three positions in $x$, two in $y$) that produce a filled "tile" covering $\sim$1.4$ \times 1.1^{\circ}$, with most positions observed twice due to the substantial overlap in the $x$ direction. However, the six pawprints must be treated separately for precise astrometric work.
VISTA and VIRCAM are described in great detail by \citet{sutherland15}. Pipeline data reduction, catalogue generation and calibration of the photometry and astrometry are provided by the Cambridge Astronomical Survey Unit (CASU), see  \citet{lewis10} and \href{http://casu.ast.cam.ac.uk/surveys-projects/vista/technical}{casu.ast.cam.ac.uk/surveys-projects/vista/technical}. The VSA provides further processing (band-merging, production of light curves etc.) and curation of VVV data and makes it available to the community as an SQL database, providing an alternative to the ESO Archive.

The raw VVV FITS file catalogues were processed by a modified version of a fortran routine \textsc{fitsio\_cat\_list} (originally provided by CASU) ported to python. This modified version unpacks the binary tables for each extension in the FITS file and calculates calibrated magnitudes from the fluxes. It also flags and computes approximate magnitudes for saturated sources using a ring-shaped aperture, removes columns which are surplus to our requirements and outputs the resultant tables as a single extension FITS file.

The flux/magnitude aperture size we selected was aperMag2 ($\text{radius}=1/\sqrt{2}\times{}1$\arcsec). This relatively small aperture produces more reliable magnitudes in crowded fields \citep{lucas08} than the more commonly used aperMag3, and aperMag2 benefits from more precise aperture corrections than aperMag1 ($\text{radius}=1/2\times{}1$\arcsec). Note the typical full width at half maximum seeing for VVV observations is $0.75$\arcsec.

For observation quality evaluation we stripped a subset of the header information from each catalogue, including airmass, seeing and the source counts for each chip. We compute the seeing for each pawprint as the median of the individual array seeing values multiplied by their plate scale (calculated from the astrometric fit coefficients of each array).

Coincident pawprints (pawprint sets) were identified by matching the telescope pointing coordinates of all pawprints using an internal sky match with a 20" matching radius with the \textsc{topcat} software package \citep{taylor05}. This yielded 2100 pawprint sets which corresponds to 6 pawprint sets for each of 346 VVV tiles and 12 pawprint sets for each of 2 VVV tiles (d015 and b390) for which the telescope pointing positions were $>1$' from their usual positions at a number of epochs. These two tiles were subject to a change in pointing coordinates due to guiding problems caused by non-stellar profiles of guide stars used by the telescope guiding system. Because of this, observations before and after the change in pointing coordinates are treated separately by our astrometric pipeline, until the final stage of averaging over the independent astrometric solutions.

We rejected pawprints based on the following criteria:
\begin{itemize}
  \item Deprecated (i.e. flagged as poor quality) by the quality control procedures used in the public data releases available at the VSA. 
  \item Seeing $>1.2$\arcsec.
  \item One or more of the 16 arrays contained fewer than 25\% of the median source counts, computed for all spatially coincident arrays not already rejected.
  \item The median r.m.s. astrometric residual of reference stars used in the CASU pipeline global astrometric solution (FITS keyword: STDCRMS) across all arrays is greater than $0.2$\arcsec.
  \item The median average stellar ellipticity (FITS keyword: ELLIPTIC) across all arrays is greater than $0.2$.
\end{itemize}

\section{Astrometric Method}\label{astrometry}

\subsection{Source Matching}

Many factors need to be considered when devising a suitable matching strategy across the many epochs of data for each pawprint set. With many epochs we are not limited by the quality of the worst epoch but we need to consider that not all sources will be detected at every epoch.

We settled on a strategy of identifying groups of epochs separated from other groups by at least 90 days, then identifying a primary epoch in each group (which we designate the P2 epoch). Most often the separations between groups correspond to the separations between observing seasons. The P2 epoch in each group is the observation with the best seeing that also has higher than median source counts for observations in the group.
The remaining epochs we refer to as secondary epochs.
The P2 epochs from consecutive groups are then matched with a 1\arcsec~OR match using the \textsc{STILTS} software package\footnote{We found that when performing an internal match of crowded UKIDSS or VISTA catalogues a 1\arcsec~matching radius typically returned only self-matches whereas $>1$\arcsec matching radii returned significant numbers of additional matches.} \citep{taylor06}. The secondary epochs are matched to the closest chronological P2 epoch with a 1\arcsec match in \textsc{STILTS} such that all P2 epoch catalogue rows are returned matched to the closest secondary epoch row. Each secondary epoch row is matched only once and unmatched P2 epoch rows are also retained.

\begin{figure}
  \begin{center}
    \begin{tabular}{c}
      \epsfig{file=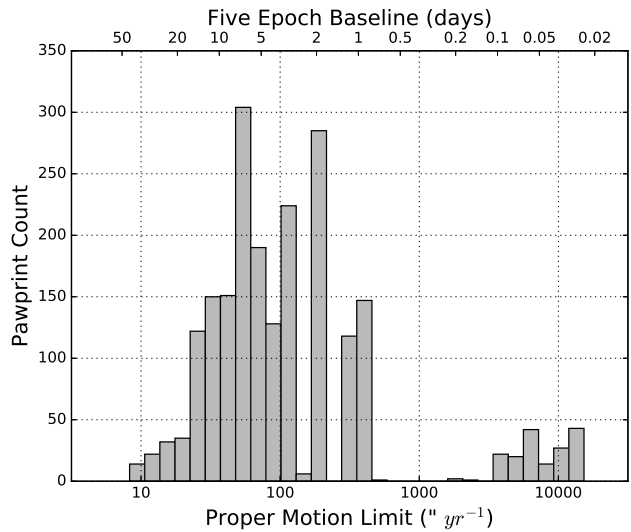,width=1\linewidth,clip=}
    \end{tabular}
    \caption{A histogram of the number of pawprints versus their maximum proper motion detection limit. The maximum proper motion detection limit comes from our 1\arcsec matching radius and our requirement that a source be detected in one P2 epoch and at least four other epochs. The equivalent epoch baseline in days is shown on the upper $x$ axis.}
    \label{pmlimits}
  \end{center}
\end{figure}

The main strengths of this matching process are as follows. For a source to be retained through matching it need only be detected in one of the P2 epochs. Clearly, for a proper motion to be measured we must have a second detection but this may come from any other P2 or secondary epoch. This means the theoretical maximum proper motion detection limit is constrained by the 1\arcsec~matching radius and the shortest epoch baseline between any P2 epoch and any other epoch. We later reject all sources detected in fewer than 5 epochs, so in reality, our maximum proper motion detection limit is constrained by the shortest epoch baseline between any P2 epoch and its fourth closest additional epoch. 
For 99.7\% of pawprints this value is less than 36.5 days, which is equivalent to a maximum proper motion detection limit of $10\arcsec~yr^{-1}$ or greater with our 1\arcsec~matching radius (see Figure \ref{pmlimits}). This is sufficient to include any very nearby stars or brown dwarfs in the VVV area.

\subsection{Coordinate Transformation}
For each pawprint set we select the observation with the best seeing as the overall primary epoch (which we designate the P1 epoch), to this we fit the remaining epochs. We perform the fit by selecting an initial pool of astrometric reference sources which meet the following criteria in the P1 epoch:
\begin{itemize}
  \item $12.5<K_s<16.0$
  \item classified as stellar
  \item $\sigma_{K_s}<0.15$ mag
  \item ellipticity less than 0.3
  \item Location $> 6$ pixels from the edge of the array in both the $x$ and $y$ dimensions
\end{itemize}

We split each array into a $5\times{}5$ grid of sub-arrays (25 sub-arrays per array, 400 per pawprint), the coordinate fitting and transformation is performed on this smaller scale. The purpose of this is to aid in correcting smaller scale non-uniformity in the focal plane (see e.g. \citet{libralato15}). The coordinate transformation fitting also incorporates astrometric reference sources from a 20 pixel wide boundary outside the sub-array. This boundary helps by providing some additional reference sources in more sparse regions. While this is not necessary in the majority of the VVV, which has very high source densities, we prefer to treat the entire survey in a homogeneous manner. Additionally, the 20 pixel boundary reduces potential edge-effects, though these should be minimal with the straightforward linear fit we use.

We then fit a linear transformation of array coordinates ($x$,$y$) for each epoch to its P1 epoch using \textsc{curve\_fit}\footnote{\label{curve_fit_note}\textsc{curve\_fit} is part of the optimization and root finding sub-module \textsc{optimize}, itself a part of the open-source mathematics, science, and engineering python module \textsc{SciPy}, see \href{https://docs.scipy.org/doc/scipy-0.17.0/reference/generated/scipy.optimize.curve_fit.html}{docs.scipy.org/doc/scipy-0.17.0/reference/generated/scipy.optimize.curve\_fit.html}}. We apply the computed transformation and remove sources with residuals greater than $3\sigma$ from the reference source pool. We repeat the fitting-transformation-rejection procedure until the reference source set does not change, leaving us with our final transformed coordinates.

\subsection{Proper Motion and Parallax Fitting}
For a proper motion measurement, the array coordinates are fit against epoch using a robust non-linear least squares method provided by \textsc{curve\_fit}. For the robust aspect, we used the Trust Region Reflective algorithm with an \textit{arctan} loss function and the default \textit{f\_scale} parameter. Testing of the different loss functions and \textit{f\_scale} parameters on various VVV tiles indicated that while there was little difference between them, the \textit{arctan} loss function produced marginally lower statistical uncertainties while still producing consistent results between the overlapping areas of adjacent pawprint sets. Comparison with an unweighted least squares fit shows that this robust method usually delivers essentially the same result as the latter. The main effect is to improve results for sources having unusually large residuals to the fit. The fit on each axis produces a proper motion, uncertainty and the epoch 2012.0 position in the P1 array coordinate frame. We experimented with calculation of a $\chi^2_{red}$ goodness of fit statistic for every solution but we found that the positional uncertainties at each epoch were not sufficiently well defined to do this accurately, especially for bright stars.

We perform a zenith polynomial projection of these positions, proper motions and uncertainties in the P1 epoch array coordinate frame using the astrometric parameters contained in the header of the original FITS file catalogue of the P1 epoch.

We fit parallaxes only for sources with proper motion greater than 20~mas~yr$^{-1}$ and detections at more than 10 epochs. We first perform the zenith polynomial projection of all positions and uncertainties as before and then fit their $\chi$ and $\eta$ positions in the tangent plane to the parallax equations in the two dimensions using the same procedure and parameters as for the proper motion fits. This produces parallaxes, proper motions and epoch 2012.0 positions in both dimensions.

It's important to note that for the moment we have not corrected for the average motion of the astrometric reference sources used. All proper motions and parallaxes are therefore relative to the average motion of sources within a few arcminutes. While an approximate relative to absolute correction can be made using a Galactic population model or by measuring motions of galaxies in
some bulge fields (see later) this cannot be done uniformly across the survey without introducing substantial uncertainty, due to the lack of precise 3D extinction maps in the Galactic plane. Such information is needed in order be confident about the median distances and motions of stars used as astrometric references.

We find in Section \ref{selfconsistency} that the difference in average motion of partially overlapping reference frames is essentially indistinguishable. Care must be taken however, if one wishes to use these data to e.g. investigate Galactic motion across larger scales (see Section \ref{galrotcurve}). For practical purposes, the changes in the astrometric reference frame can typically be neglected on scales of approximately a VVV tile, though there may well be exceptions to this in mid-plane fields with highly structured extinction (which affects the distance and motions of the reference stars).

\section{Proper Motion Results}\label{pmresults}

\subsection{Self-Consistency}\label{selfconsistency}
The VVV observation method enables a check for self-consistency between overlapping pawprint sets. Overlaps on the sky between pawprint sets are either between different arrays or different sections of the same array, in different parts of the focal plane. Since the volume of data is so great we select a sample of sources from tiles b216 (outer bulge), b332 (inner bulge, very high source density), d069 (inner disk, containing the Westerlund 1 compact young cluster) and d079 (outer disk) to test self-consistency. 
On these tiles we perform a 1\arcsec~internal match on their 2012.0 positions to identify coincident detections of sources between multiple pawprint sets. We compare proper motions for sources with solutions in two pawprint sets and no proper motion error flags (see Section \ref{epm_flagging}) and find that in each case the random errors in the proper motion measurements are described well by Gaussian distributions with the statistical uncertainties provided by \textsc{curve\_fit} (see Table \ref{sctable}).

\begin{table}
\centering
\caption{Fractions of proper motions which agree between overlapping pawprints of four VVV tiles to within $1\sigma$, $2\sigma$, and $3\sigma$ uncertainties.}
\label{sctable}
\begin{tabular}{|l|l|c|c|c|}
\hline
  \multicolumn{1}{|c|}{tile} &
  \multicolumn{1}{c|}{dimension} &
  \multicolumn{1}{c|}{$1\sigma$} &
  \multicolumn{1}{c|}{$2\sigma$} &
  \multicolumn{1}{c|}{$3\sigma$} \\
\hline
 b216 & $\alpha\cos\delta$ & 0.691 & 0.955 & 0.997 \\
      &           $\delta$ & 0.683 & 0.951 & 0.996 \\
 b332 & $\alpha\cos\delta$ & 0.682 & 0.945 & 0.994 \\
      &           $\delta$ & 0.683 & 0.944 & 0.994 \\
 d069 & $\alpha\cos\delta$ & 0.680 & 0.945 & 0.995 \\
      &           $\delta$ & 0.686 & 0.948 & 0.995 \\
 d079 & $\alpha\cos\delta$ & 0.689 & 0.954 & 0.997 \\
      &           $\delta$ & 0.684 & 0.951 & 0.996 \\
\hline\end{tabular}
\end{table}

\subsection{Proper Motion Quality Flagging}\label{epm_flagging}

\begin{figure}
  \begin{center}
    \begin{tabular}{c}
      \epsfig{file=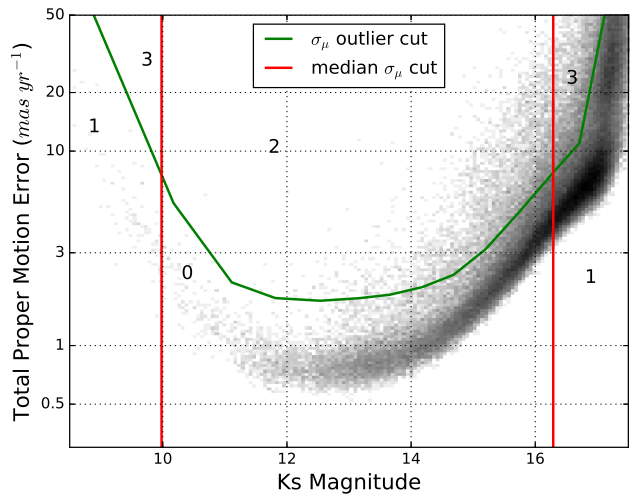,width=1\linewidth,clip=}
    \end{tabular}
    \caption{A 2d histogram showing proper motion uncertainty vs. magnitude for one pawprint set of tile b216 before averaging of measurements from overlapping pawprint sets. Note the rapid increase in proper motion measurement errors at the bright and faint limits, and the presence of some sources in the middle of the plot with unusually large proper motion uncertainties for their magnitude. The red lines show the magnitude range outside which we flag poorly measured sources (epm flag=1). The green line is the trace above which we flag sources as proper motion uncertainty outliers (epm flag=2). The numbers indicate the epm flag assigned to each region (the overlap of epm flag=2 and epm flag=1 gives epm flag=3).}
    \label{epm_flag}
  \end{center}
\end{figure}

\begin{figure}
  \begin{center}
    \begin{tabular}{c}
      \epsfig{file=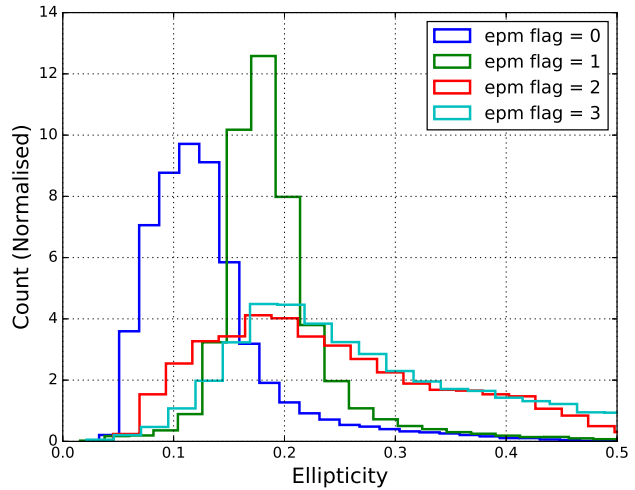,width=1\linewidth,clip=}
    \end{tabular}
    \caption{The ellipticity distributions of the separate resultant groups from the epm flagging routine of a single pawprint set of tile b216. Sources with an epm flag greater than 0 tend to have a higher ellipticity (suggestive of blending). Only sources with {\it epm flag} = 0 in all pawprints sets, and solutions in at least two pawprint sets, are given the {\it reliable} flag, see Section \ref{pmcat}.}
    \label{epm_flag_groups}
  \end{center}
\end{figure}

Figure \ref{epm_flag} shows the magnitude vs. proper motion uncertainty range for one pawprint set of VVV tile b216. Note the rapid increase in proper motion uncertainties at the bright and faint end, and the spread due to some sources in the middle with uncharacteristically large proper motion uncertainties for their magnitudes. For individual pawprint sets, we use a proper motion error flag, \textit{epm flag}, that identifies these regions, in which reliability of proper motion measurements is generally low. This is most often a result of saturation of bright stars, low signal to noise ratio for faint stars, or most notably blending, as evidenced by relatively high ellipticity of sources in these regions (see Figure \ref{epm_flag_groups}).

After averaging the multiple solutions for each source (see Section \ref{pmcat}) the \textit{epm flag} information is used to set a simple \textit{reliable} flag (1 is reliable, 0 is not) to facilitate selection of the most reliable VIRAC proper motion measurements. 

To define \textit{epm flag} for each pawprint set, we group sources by $K_s$ magnitude in bins with 500 sources. Bins narrower than 0.5~mag are joined with neighbouring bins. We interpolate over the median proper motion uncertainties vs. median magnitudes for each group and identify the bright and faint limits outside which the median proper motion uncertainty exceeds 5~mas~yr$^{-1}$. These represent the points at which the reliability of the results significantly decreases and such results are given \textit{epm flag}=1.
We also identify sources in each bin with proper motion uncertainties larger than the median proper motion uncertainty plus three times the spread, defined as the larger of 0.3~mas~yr$^{-1}$ and the median absolute deviation. 
Sources with proper motion uncertainty above this threshold, determined by interpolation of the thresholds across adjacent bins are given \textit{epm flag}=2. The 0.3~mas floor is imposed in order to avoid flagging sources with errors that are no more than about double the typical error in the magnitude bin.
In cases where both conditions are true we give sources \textit{epm flag}=3. Where no flags are set \textit{epm flag}=0.
Figure \ref{epm_flag} illustrates these selections made on one pawprint set of tile b216.

\subsection{The Catalogue: Averaged Proper Motions}\label{pmcat}

The tile and pawprint pattern of the survey is such that, with the exception of the very edges of the survey, all sources should be observed in at least two pawprint sets and therefore have a proper motion measurement in each. Additionally, due to the matching method we employ, some sources (e.g. faint, or very high proper motion stars) are not matched between consecutive P2 epochs and hence will also have a proper motion measured for multiple epoch groups within each pawprint set. To produce a catalogue of unique sources we identify groups of proper motion measurements by matching epoch 2012.0 positions to within 1\arcsec, and average their proper motions using inverse variance weighting. This matching and averaging is performed across tiles as well as pawprint sets and VIRAC proper motions are split into tile catalogues (d001 to d152, b201 to b296) with sources common to two or more tiles usually assigned to the catalogue with the smallest tile number.

For each proper motion measurement, we report a $K_s$ magnitude and uncertainty as the median and median absolute deviation (respectively) across all epochs that go into the proper motion measurement. For a source morphological classification we report the modal classification across those epochs. When we come to combine multiple measurements we give the inverse variance weighted average $K_s$ magnitude and simply the number of proper motion measurements that have a modal stellar classification. The epm flags applied to each proper motion measurement are retained as counts, these are incorporated into a simple '\textit{reliable}' flag. To be flagged as '\textit{reliable}' a source must have a minimum of two proper motion measurements, all proper motion measurements must be from different pawprint sets, and there must be no error flags set for any proper motion measurement.

\begin{figure*}
  \begin{center}
    \begin{tabular}{c}
      \epsfig{file=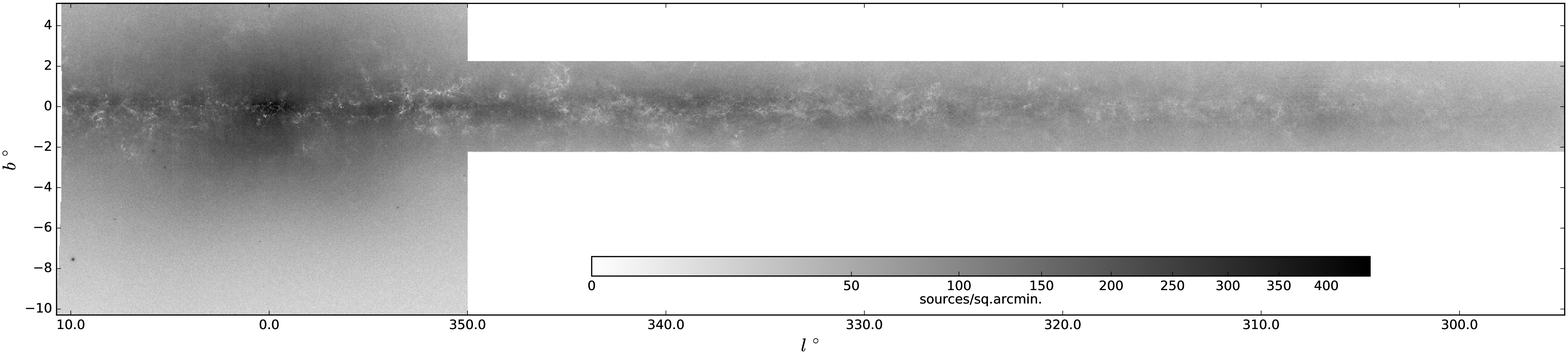,width=1\linewidth,clip=}\\
      \epsfig{file=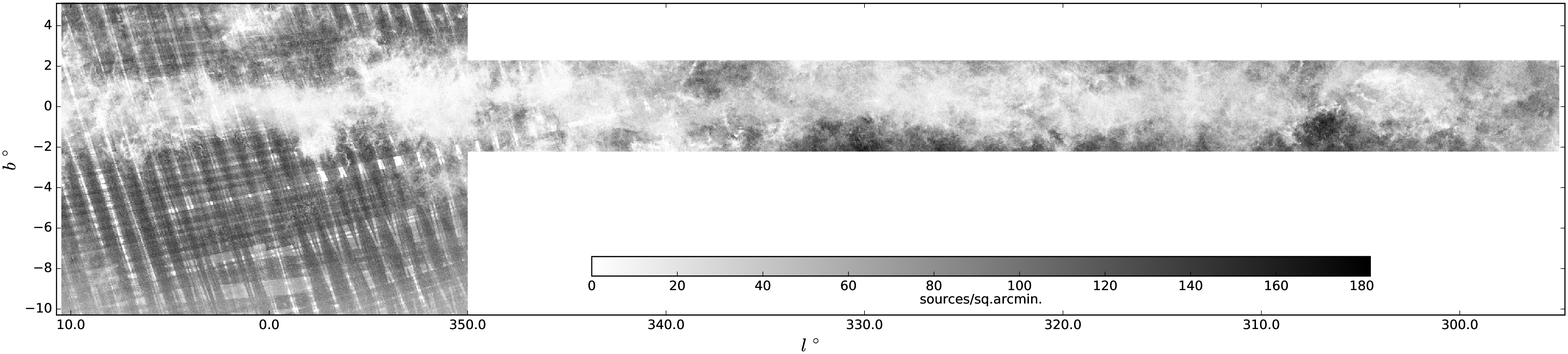,width=1\linewidth,clip=}\\
      \epsfig{file=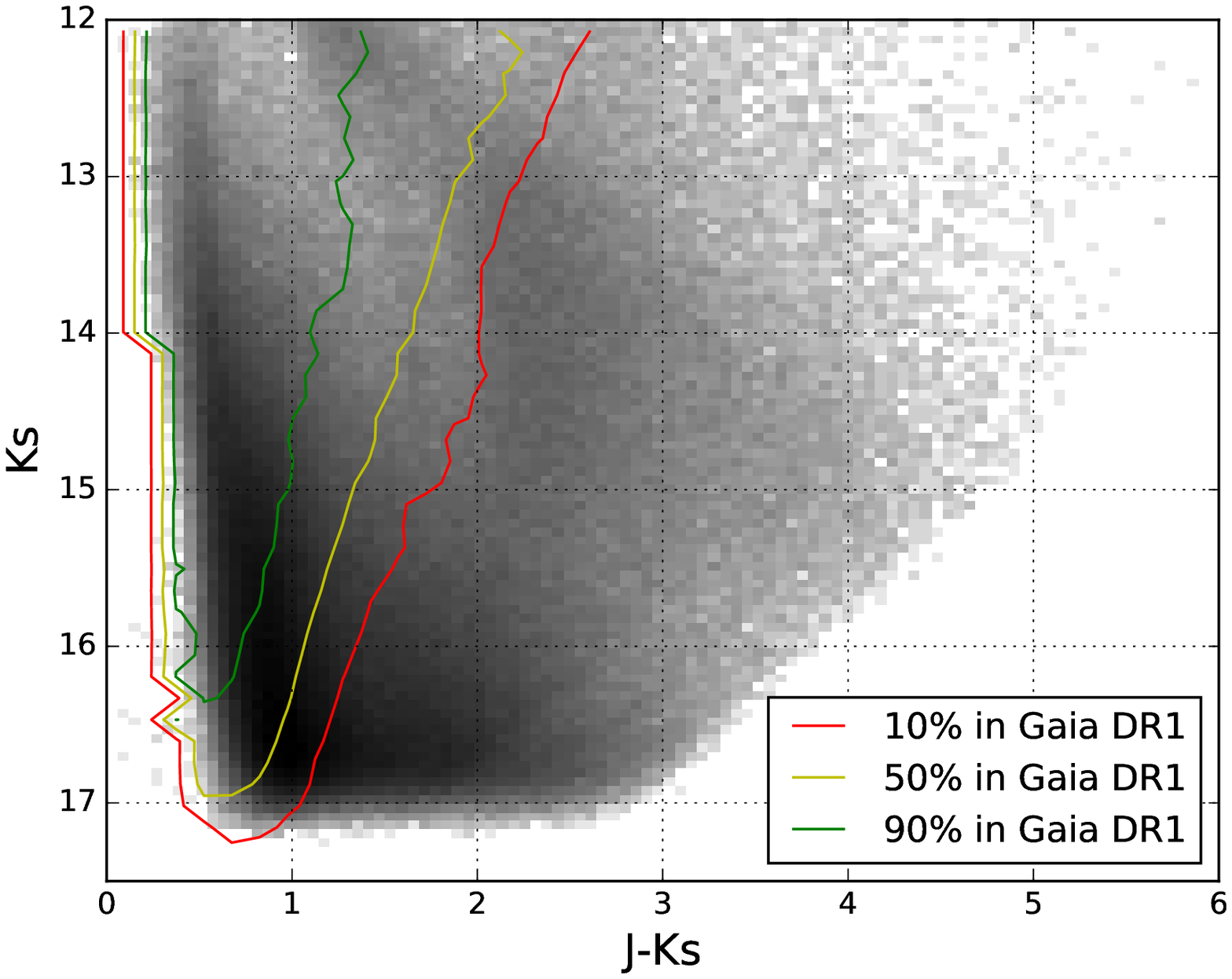,width=0.5\linewidth,clip=}
    \end{tabular}
    \caption{\textit{upper}: Our VVV proper motion catalogue area coverage, with greyscale showing the $10<K_s<16$ source density. \textit{middle}: The Gaia DR1 source density in the VVV survey area for comparison. \textit{lower}: The K$_{s}$ versus J-K$_{s}$ colour-magnitude diagram of VIRAC sources in VVV tile d084 ($l\approx{}305.67^{\circ}$, $b\approx{}0.52^{\circ}$). We crossmatched these against the Gaia DR1 catalogue, the contours show the regions inside which 10\%, 50\%, and 90\% of VIRAC sources were detected by Gaia.}
    \label{vvv_source_density}
  \end{center}
\end{figure*}

\begin{figure}
  \begin{center}
    \begin{tabular}{c}
      \epsfig{file=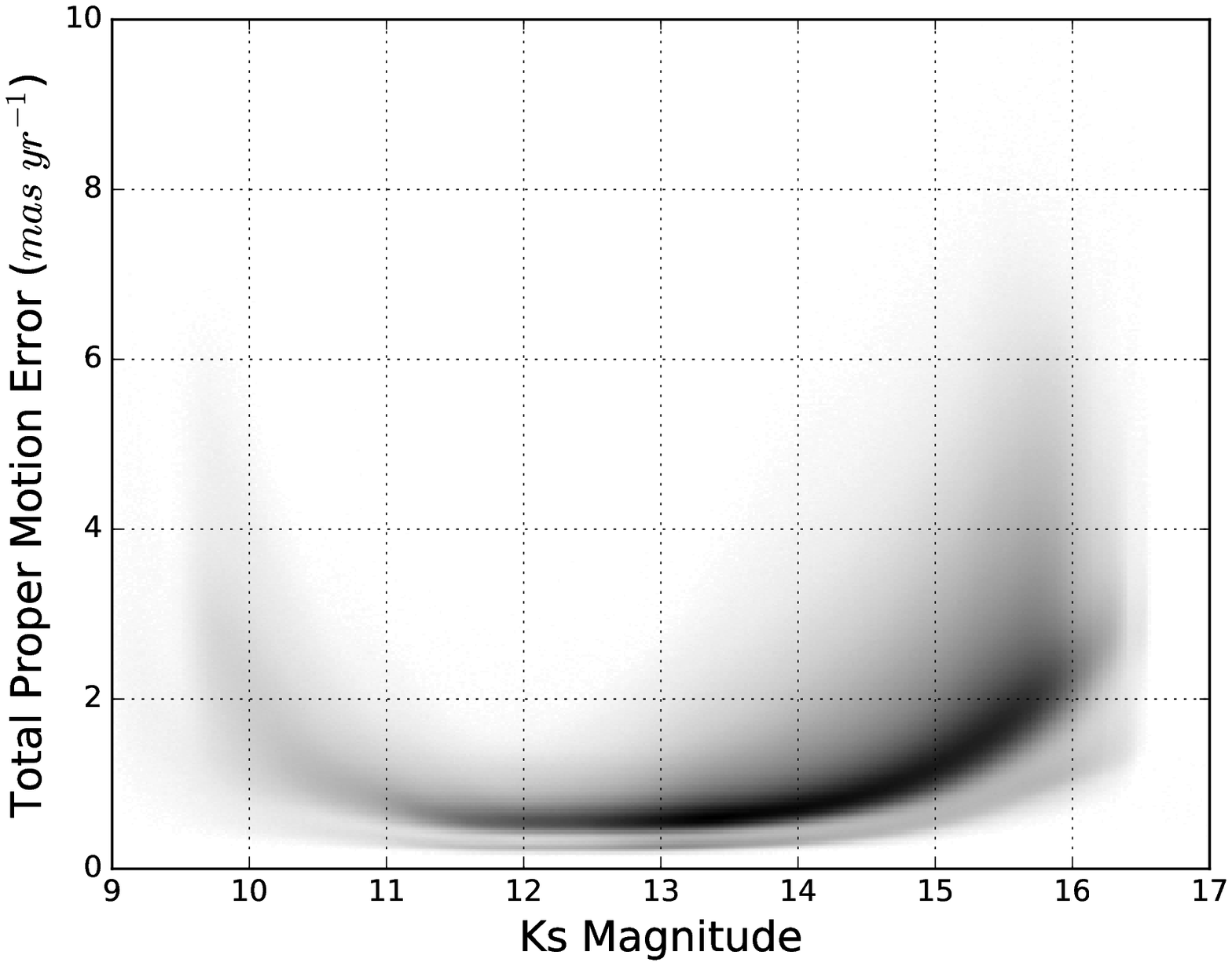,width=1\linewidth,clip=}
    \end{tabular}
    \caption{Density plot of $K_s$ magnitude vs. uncertainty on the total proper motion of sources with detections in multiple pawprints and flagged as {\it reliable} (see Section \ref{epm_flagging}). We consider this to be a reasonably reliable selection of proper motion measurements encompassing 119 million sources, 47 million of which have sub-1~mas~yr$^{-1}$ proper motion uncertainties. The track of sources visible underneath the main body is formed by 8 high cadence bulge tiles with several hundred epochs and very precise proper motions.}
    \label{mag_epm}
  \end{center}
\end{figure}

The combination of multiple proper motion measurements as above yielded 312,587,642 unique sources. Figure \ref{vvv_source_density} shows the area covered and the source density and compares this to the Gaia DR1. Figure \ref{mag_epm} shows proper motion uncertainty vs. magnitude distribution of the 119 million sources we consider to have the most reliable proper motions.

\subsection{Comparison to a Visually Confirmed Proper Motion Sample}\label{kurtev_comparison}

\begin{figure}
  \begin{center}
    \begin{tabular}{c}
      \epsfig{file=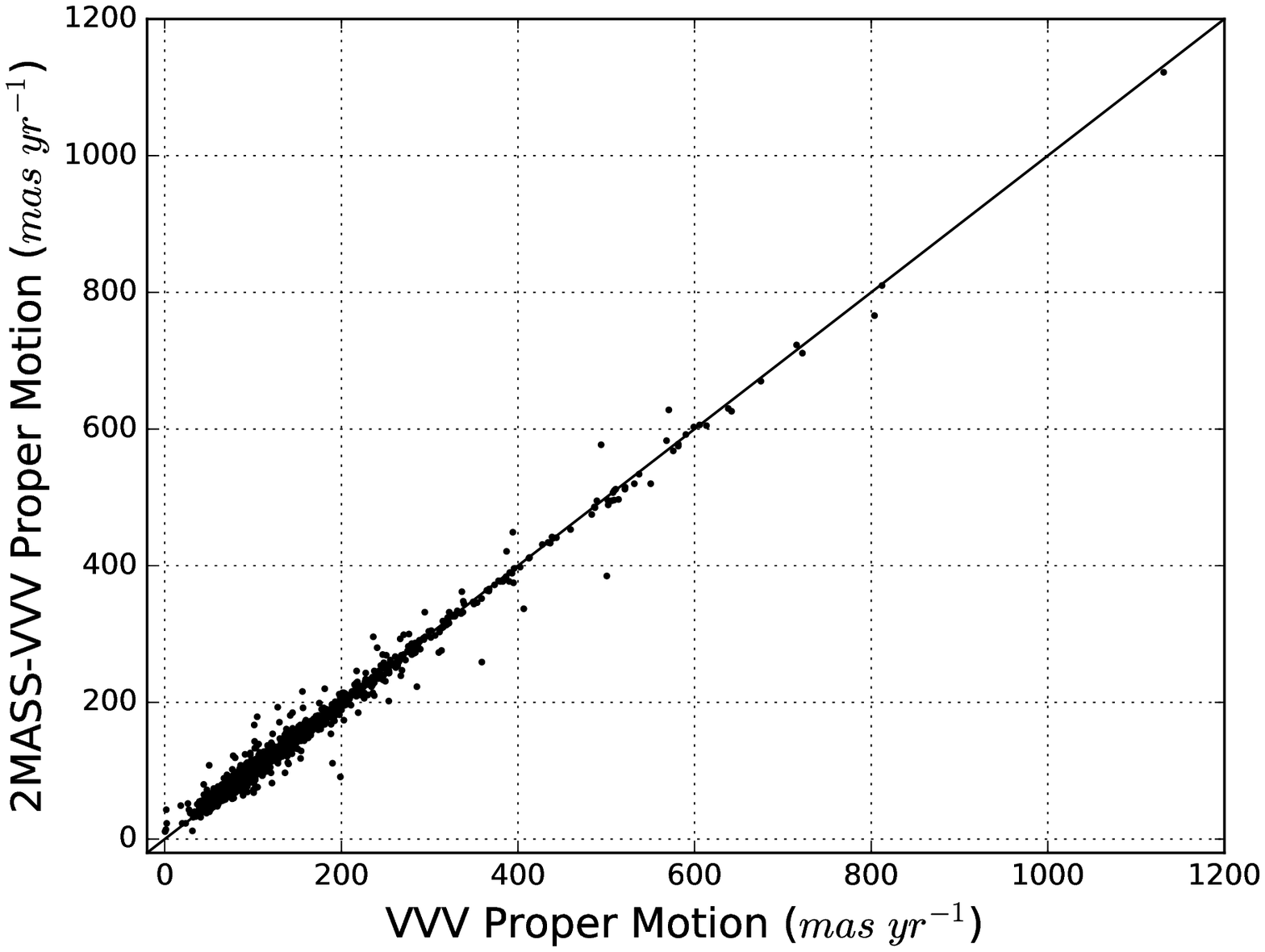,width=1\linewidth,clip=}
    \end{tabular}
    \caption{Our VVV total proper motions vs. the 2MASS $+$ VVV total proper motions of \citet{kurtev17} for the 2847 sources recovered from their high proper motion catalogue of bright stars.}
    \label{kurtevpmcompare}
  \end{center}
\end{figure}

\citet{kurtev17} produced a visually confirmed sample of 3003 proper motion sources in the VVV area. For most sources they provide a 2MASS to VVV proper motion solution, covering a time baseline between 10 and 15 years. Their sample covers proper motions typically in the range 50--1000~mas~yr$^{-1}$, and magnitudes from the brightest end of the VVV survey to $K_s \approx 13.5$.
We removed $\eta$ Sagittarii (source 2679) from the \citet{kurtev17} list as it is far too bright for any meaningful VVV detection (2MASS $K_s=-1.55$) and a further 151 sources for which proper motion was not given in their catalogue because their inclusion was based on previous proper motion measurements. Among the remaining sources we removed a further 4 duplicates: 1064/1065, 1134/1135, 1892/1893, 2394/2395. This left a total of 2847 sources which we should be able to recover in VIRAC. On crossmatching this list to our full results table (i.e. allowing sources not flagged as reliable) we recover all 2847 objects. Figure \ref{kurtevpmcompare} shows the \citet{kurtev17} 2MASS-VVV total proper motion versus those of VIRAC for these objects.

\begin{figure}
  \begin{center}
    \begin{tabular}{c}
      \epsfig{file=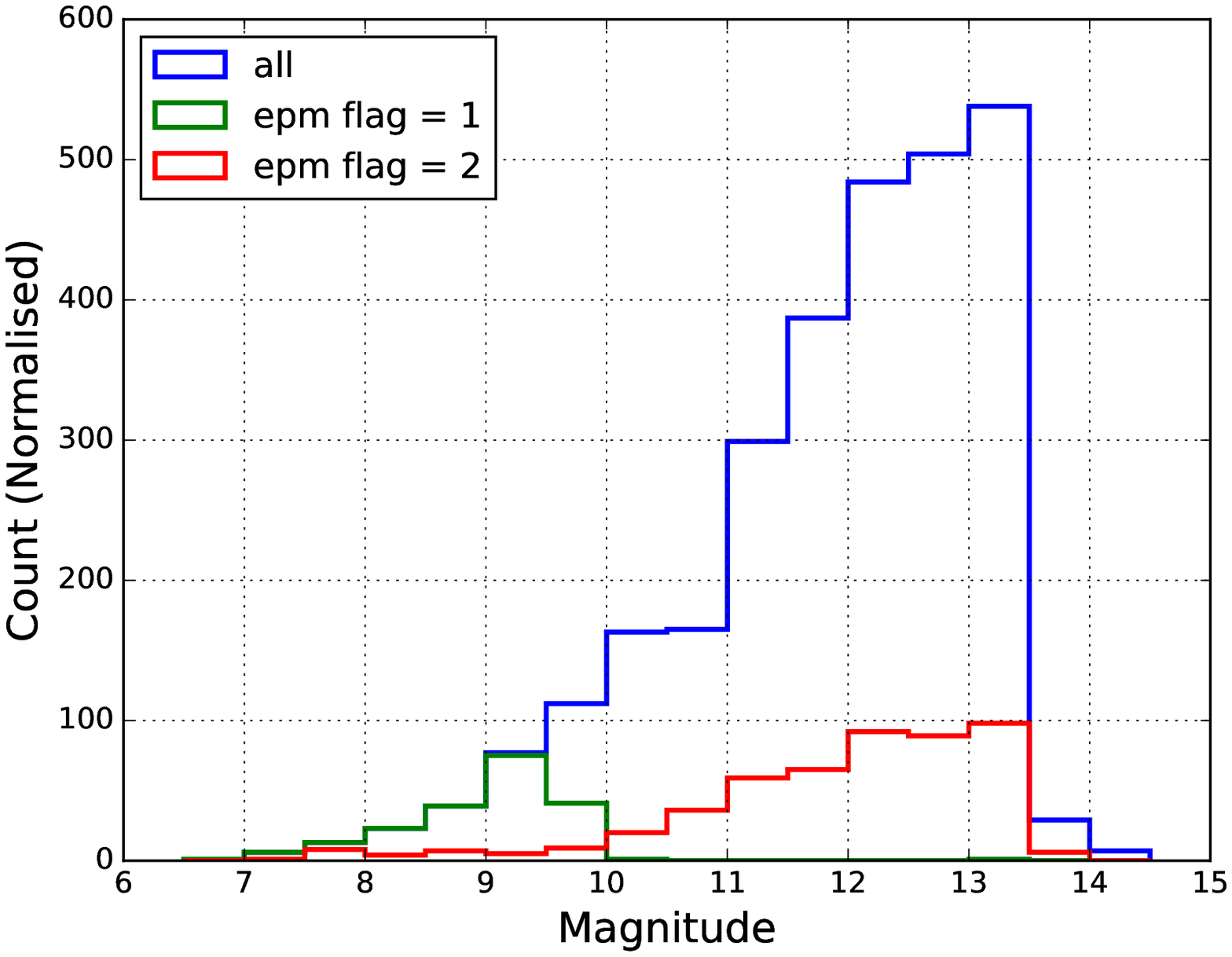,width=1\linewidth,clip=}
    \end{tabular}
    \caption{The breakdown of our epm flags by magnitude for known genuine high proper motion sources in common with the \citet{kurtev17} search. A source need only have an {\it epm flag} for a single proper motion measurement to be included in that histogram. Note that {\it epm flag}=3 indicates both {\it epm flags} 1 and 2 for a proper motion measurement so we have included such cases in both epm flag histograms. The blue histogram shows the breakdown of all 2847 sources by magnitude.}
    \label{kurtevepmfcompare}
  \end{center}
\end{figure}

We used this sample of 2847 bona fide high proper motion stars to evaluate our {\it epm flag} system, at least for the brighter end of our results. Figure \ref{kurtevepmfcompare} shows the breakdown of sources with any {\it epm flag} by magnitude. As one might expect, sources at the very brightest end of the survey have an {\it epm flag} indicating that their magnitude bin has median proper motion uncertainty greater than 5~mas~yr$^{-1}$ ({\it epm flag}=1) consistent with them saturating. The presence of many sources with an {\it epm flag}=2, meaning their proper motion uncertainties are significantly higher than normal for their magnitude is interesting. We visually inspected a sample of these sources ourselves and found that overwhelmingly the reason for their higher than normal proper motion uncertainties was blending with a background source. Another reason for {\it epm flag}=2 to be set for high proper motion sources is that many will have significant parallactic motion that would cause large scatter about a linear fit of position vs time (i.e. proper motion alone).
This suggests that if one is looking for a \textit{complete} selection of sources it is advisable to ignore our source reliability flag and include sources with {\it epm flag}=2.

\subsection{Visually Confirmed High Proper Motion Stars}\label{visinspect}

\begin{figure}
  \begin{center}
    \begin{tabular}{c}
      \epsfig{file=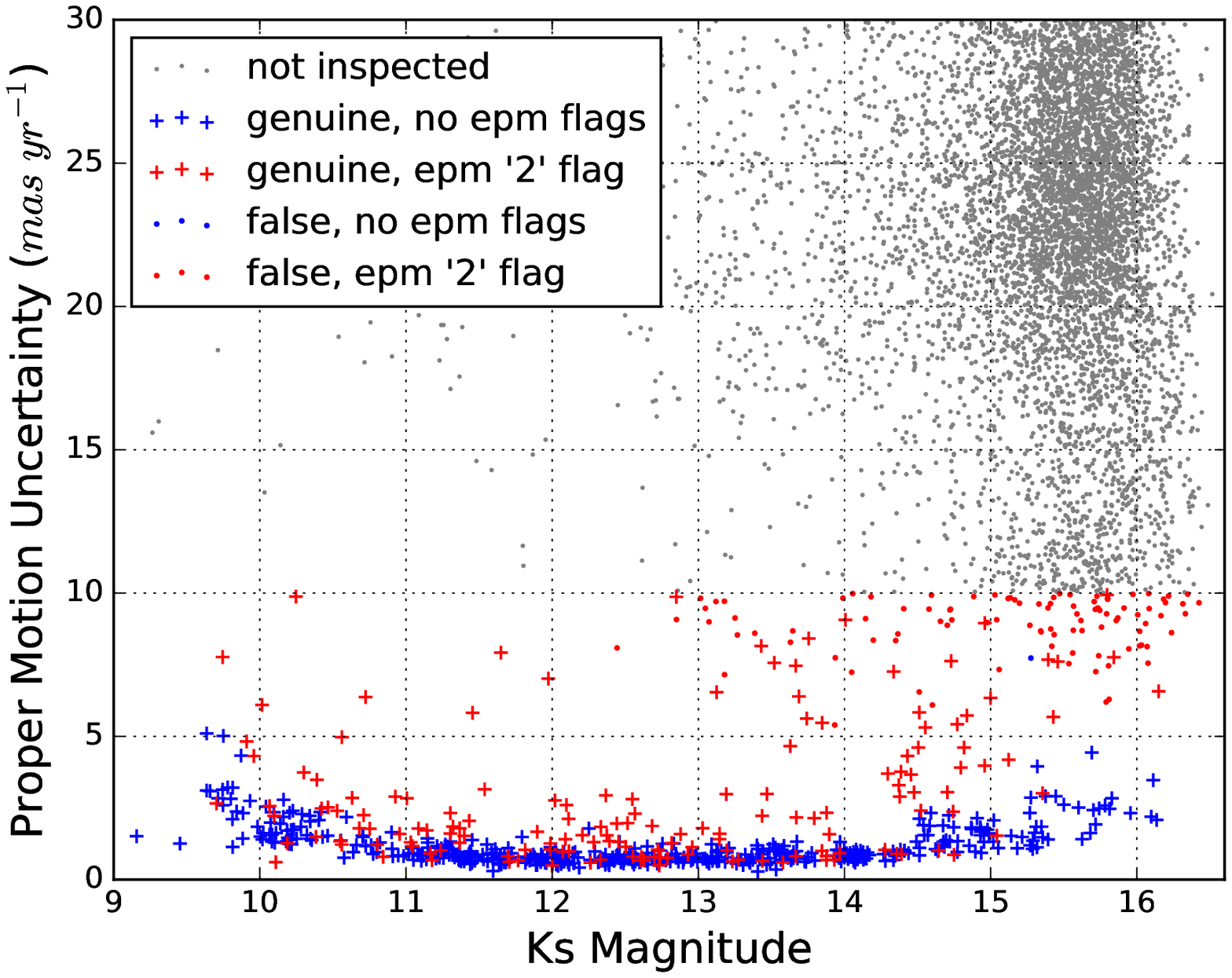,width=1\linewidth,clip=}
    \end{tabular}
    \caption{Proper motion error vs magnitude for the 6,796 sources with $\mu > 200$~mas~yr$^{-1}$ and $\sigma_{\mu} < 30$~mas~yr$^{-1}$. Sources with $\sigma_{\mu} < 10$~mas~yr$^{-1}$ were visually inspected. Genuine high proper motion sources are identified by pluses, false high proper motion sources by blue or red dots. Blue points are those flagged as 'reliable', red points allow {\it epm flag} = 2.}
    \label{visinspect_sample}
  \end{center}
\end{figure}

To assess the reliability of high proper motion source detections we selected sources with at least two proper motion measurements, proper motion, $\mu$, measured with at least $5\sigma$ significance and $\mu > 200$~mas~yr$^{-1}$. We rejected the very brightest and very faintest sources (sources with an {\it epm flag} = 1 or 3 in any proper motion measurement, see Section \ref{epm_flagging}) but we retain sources with {\it epm flag} = 2 in one or more solutions (hence not flagged as {\it reliable}) so as to include sources that might otherwise be excluded due to parallactic motion or blending. This yielded 14,921 sources. Their proper motion uncertainty vs. magnitude distribution is shown in Figure \ref{visinspect_sample} for the 6,796 with $\sigma_{\mu}<30$~mas~yr$^{-1}$. We visually inspected the 687 sources in this sample with $\sigma_{\mu}<10$~mas~yr$^{-1}$, of which 255 have {\it epm flag}=2 and would be expected to be less reliable. The 687 sources are identified in Figure \ref{visinspect_sample} by coloured pluses and dots for those identified as genuine and false respectively.

All of the 432 "reliable" sources with {\it epm flag} = 0 were visually confirmed as genuine, save for one ambiguous case. This shows that this citerion is indeed useful for making a reliable selection. The ambiguous source emerged from a blend with a background source over the 2010-2015 period: it appears to have a genuine motion but could perhaps be explained as a gradually brightening variable star.

\begin{figure}
  \begin{center}
    \begin{tabular}{c}
      \epsfig{file=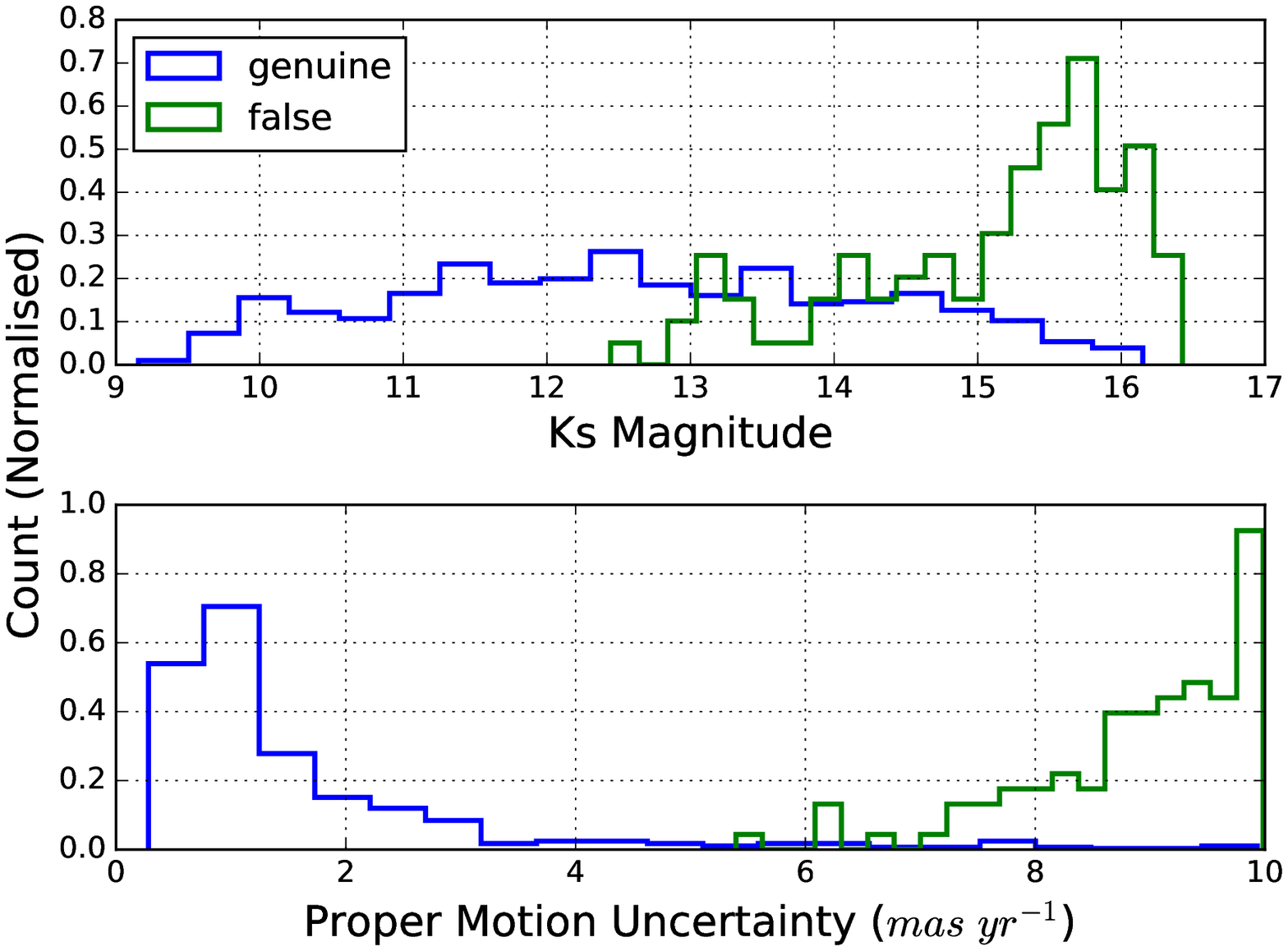,width=1\linewidth,clip=}
    \end{tabular}
    \caption{\textit{upper}: The $K_s$ magnitude distribution of the sample of the 687 sources with $\mu>200$~mas~yr$^{-1}$ and $\sigma_{\mu}<10$~mas~yr$^{-1}$ that we visually inspected. \textit{lower}: The proper motion uncertainty of the same sample. In both cases the blue histogram shows the distribution of objects we've verified are genuine.}
    \label{visinspect_hist}
  \end{center}
\end{figure}

Figure \ref{visinspect_sample} shows that the catalogue contains a locus of candidate high proper motion sources at faint magnitudes with large proper motion errors and {\it epm flag}=2 in one or more solutions (due to the large proper motion uncertainties for their magnitude bin). Our visually inspected sample grazes the bottom of this distribution of sources: we found that all sources inspected which might reasonably be considered to be part of this group had false motions, predominantly due to mismatching caused by blending. Note that a gradual shift from predominantly genuine to predominantly false occurs at $\sigma_{\mu}\approx{}5$~mas~yr$^{-1}$, see the lower panel of Figure \ref{visinspect_hist}. The presence of genuine high proper motion sources with and without {\it epm flag}=2 in Figure \ref{visinspect_sample} again highlights the need to ignore this particular flag if a complete selection is required. 

The 555 visually confirmed high proper motion sources with $\mu>200$~mas~yr$^{-1}$ and $\sigma_{\mu}<5$~mas~yr$^{-1}$ are presented in Table \ref{hpmtable}. Note that we do not remove sources we have identified as having incorrect proper motions from the full catalogue, we prefer to treat the entire catalogue in a homogeneous manner and it is impractical to visually inspect everything. Again, the 'reliable' flag can be used if one wishes to use the more reliable sample.

\subsubsection{TGAS Common Proper Motion Companions}\label{tgas_cpm}

We undertook a search for common proper motion companions to TGAS proper motion sources. Since the probability of chance alignments of unrelated sources with similar motions falls as proper motion increases we select only TGAS sources with $\mu > 100$~mas~yr$^{-1}$. We match this TGAS sample to a selection of our VIRAC sources with motions measured at $>5\sigma$ significance, allowing {\it epm flag} = 0 or 2. We find 199 matches with separation less than 300\arcsec~and proper motion consistent to within 50~mas~yr$^{-1}$ in both dimensions. To attempt to quantify the probability that each match is a chance alignment of unrelated objects we count TGAS sources in an annulus ($r_{in}=5$\arcmin, $r_{out}=300$\arcmin) around the VVV component of the match which have a proper motion agreement as good or better than that of the match. We then divide this count by the ratio of the area of the annulus and the area of a circle with radius equal to the original match separation to approximate the number of expected chance alignments within this radius. This method assumes that there will be no genuine companions outside 5\arcmin. For the closest pair at $d$=36~pc, 5\arcmin\ is a projected separation of 10800~au; known systems with separations wider than this are rare. We reject any pairs with a number of expected chance alignments greater than $10^{-4}$, and any whose projected separation at the distance of the TGAS source is greater than 10000~au. 

This left 49 promising common proper motion companion candidates. We visually inspected the VVV component of each pair and found: 11 that are TGAS detections of the VVV source (these were all candidates with separation less than 1\arcsec), 10 for which the VIRAC high proper motion is false (note that these were all cases with an {\it epm flag}=2 for one or more of their individual proper motion measurements), and 28 which have genuine high proper motion and are therefore bona-fide common proper motion companions. Table \ref{tgas_companions} lists these 28 bona-fide VVV common proper motion companions to TGAS sources.

A search of SIMBAD and the literature indicates that three of these systems are known: TYC 7365-318-1 AB was identified by \citet{kurtev17}, and L 149-77 AB and L 200-41 AB were identified by \citet{ivanov13}. Note that the \textbeta~Circini system identified by \citet{smith15} does not appear in this list since \textbeta~Circini does not appear in TGAS. The remaining 25 common proper motion pairs were previously unidentified. Of particular note are two of the systems: the CD-53 6250 AB system, a pre-main sequence K0IV primary star \citep{torres06} and an X-ray detected secondary (we believe CD-53 6250 B is source 22 of \citealt{chen08} table 1); and the LTT 7251 AB high contrast pair, a G8V primary with a low mass companion (see Section \ref{171004304}).

\subsubsection{Very high proper motion candidates}\label{addvisinspect}

In addition to the sample visually inspected in Section \ref{visinspect} we also searched for very high proper motion objects with still reasonable proper motion uncertainties, and for common proper motion companions to Proxima Centauri.

We identified four sources with $\mu > 1\arcsec~yr^{-1}$, $\sigma_{\mu}<$30~mas~yr$^{-1}$ and a minimum of two proper motion solutions that each have {\it epm flag} = 0 or 2. Of these four sources, one was genuine. See Section \ref{357154962} for details.

Our nearest stellar neighbour, high proper motion star and recently confirmed planet host \citep{anglada16} Proxima Centauri lies in the VVV survey footprint. We identified 163 sources that were within 128\arcmin\footnote{Approximately 10,000 AU at the distance of Proxima Centauri} of the \citet{vanleeuwen07} position of Proxima Centauri that also had a proper motion within $\pm~1\arcsec~yr^{-1}$ of the \citet{vanleeuwen07} proper motion for Proxima Centauri ($-3775.75\pm1.63$ and $765.54\pm2.01$ in $\mu_{\alpha\cos\delta}$ and $\mu_{\delta}$ respectively). This sample had no other selection criteria applied to it. We visually inspected the 163 sources but none had a genuine proper motion. This agrees with a recent dedicated VVV search for companions to this star by \citet{beamin17}.

\subsubsection{Reduced Proper Motion}

In lieu of parallax measurements for all sources, and since proper motion scales as the inverse of the distance (with a few exceptions), we can use reduced proper motion to estimate luminosity of objects.
Where $H_{K_{s}}$ (reduced proper motion in the $K_{s}$ band) is:

$H_{K_{s}} = K_s + 5\log{\mu} + 5$

\begin{figure}
  \begin{center}
    \begin{tabular}{c}
      \epsfig{file=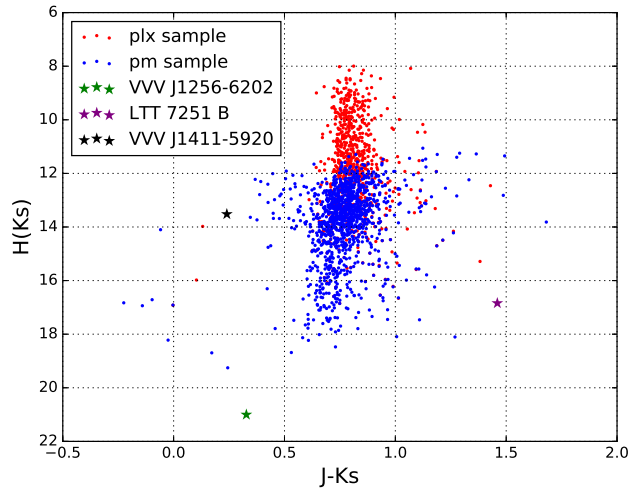,width=1\linewidth,clip=}
    \end{tabular}
    \caption{A reduced proper motion diagram of sources drawn from the 3200 visually confirmed high proper motion sources from this work and \citet{kurtev17} and the VIRAC parallax sample (see Section \ref{plxcat}). We use VIRAC proper motion, median $K_s$ magnitudes across proper motion epochs and VVV DR4 tile catalogue $J$ band photometry from the VSA. The 1167 proper motion sources in this plot comprise all those with $K_s<20$, $J<20$, $K_{s}>11$ (to minimise saturation effects), stellar $J$ band morphological classification (also removes saturated sources), $\sigma_{K_{s}}<0.07$, and $\mu > 100$~mas~yr$^{-1}$ measured at $5\sigma$ significance. The 740 parallax sources in this plot are those with $10\sigma$ parallaxes and similar photometric requirements to the proper motion selection. The low metallicity L dwarf VVV J12564352-6202041 (see Section \ref{357154962}), the L dwarf LTT 7251 B (see Section \ref{171004304}), and the nearby white dwarf VVV J14115932-59204570 (see Section \ref{solneighbours}) are also included.}
    \label{hpm_rpm}
  \end{center}
\end{figure}

Figure \ref{hpm_rpm} shows the $H_{K_s}$ vs. $K_s$ for the visually confirmed VIRAC high proper motion ($\mu>100$~mas~yr$^{-1}$) sources and including those previously identified by \citet{kurtev17}. We also include the $10\sigma$ VIRAC parallax sources (see Section \ref{plxcat}). Note that these are not restricted to $\mu>100$~mas~yr$^{-1}$ and are on average brighter than the proper motion sources, and hence have smaller values of $H_{K_{s}}$ (higher up in Figure \ref{hpm_rpm}). We also include the locations of three interesting objects: LTT 7251 B, a benchmark L dwarf companion to LTT 7251 A; VVV J12564352-6202041, a low metallicity (hence blue and low luminosity relative to LTT 7251 B) L dwarf; and VVV J14115932-59204570, a close (d$\approx$15pc) probable white dwarf with relatively low proper motion ($\mu=98.6$~mas~yr$^{-1}$).

\subsection{Further Validation of Uncertainties: NGC\,6231}\label{ngc6231}

To test the reliability of our uncertainties and investigate whether proper motions are influenced by unresolved blending we study the proper motion distribution of the \citet{kuhn17} X-ray selected members of the pre-MS cluster NGC 6231 in VVV tiles d148 and d110. VVV tiles d148 and d110 are representative of the survey in terms of source density. \citet{wilson17} showed that unresolved blending can commonly affect photometry and astrometry, and it could plausibly affect a proportion of our proper motion measurements. NGC 6231 is populous but sparse: field stars are much more numerous and typically have different proper motions to cluster members. \citet{kuhn17} estimate that 88\% of their X-ray selection with NIR counterparts are genuine cluster members. A low intrinsic velocity dispersion in the cluster translates to negligible intrinsic proper motion dispersion at the cluster distance of $\approx1.59$~kpc. 

We crossmatched VIRAC sources flagged as reliable to this sample (1193 matches) and find the average relative proper motions to be $1.88\pm0.05$ and $1.42\pm0.05$~mas~yr$^{-1}$ in $\alpha\cos\delta$ and $\delta$ respectively. If we restrict the selection to only sources with VIRAC proper motion uncertainties below $1$~mas~yr$^{-1}$ we obtain $1.89\pm0.04$ and $1.41\pm0.05$~mas~yr$^{-1}$ in $\alpha\cos\delta$ and $\delta$ respectively, a negligible difference. The proper motions of stars in the field with $\sigma_{\mu}<1$~mas~yr$^{-1}$ typically differ from this by a few ~mas~yr$^{-1}$ (median difference of 3.2~mas~yr$^{-1}$) due to a spread in motion in the Galactic longitude direction caused by Galactic rotation and their wide range of distances. By contrast, the 1193 X-ray selected cluster members show a tight distribution about the mean cluster motion in the vector point diagram, located at one end of the field distribution. The significances of the offsets from the average proper motion follow an approximate Gaussian distribution (57\% within $1\sigma$, 88\% within $2\sigma$, 97\% within $3\sigma$, these values hold for the $\sigma_{\mu}<1$~mas~yr$^{-1}$ sub-sample). When we consider that we have a contamination rate by non-cluster members of approximately 12\% this is encouraging evidence that our proper motion uncertainties are reasonably representative of the true scatter in the measurements. It also shows that unresolved blending is not a significant issue for the great majority of proper motion measurements in the sample flagged as reliable.

\section{Parallax Results}\label{plxcat}

\subsection{Quality Control Cuts}

\begin{figure}
  \begin{center}
    \begin{tabular}{c}
      \epsfig{file=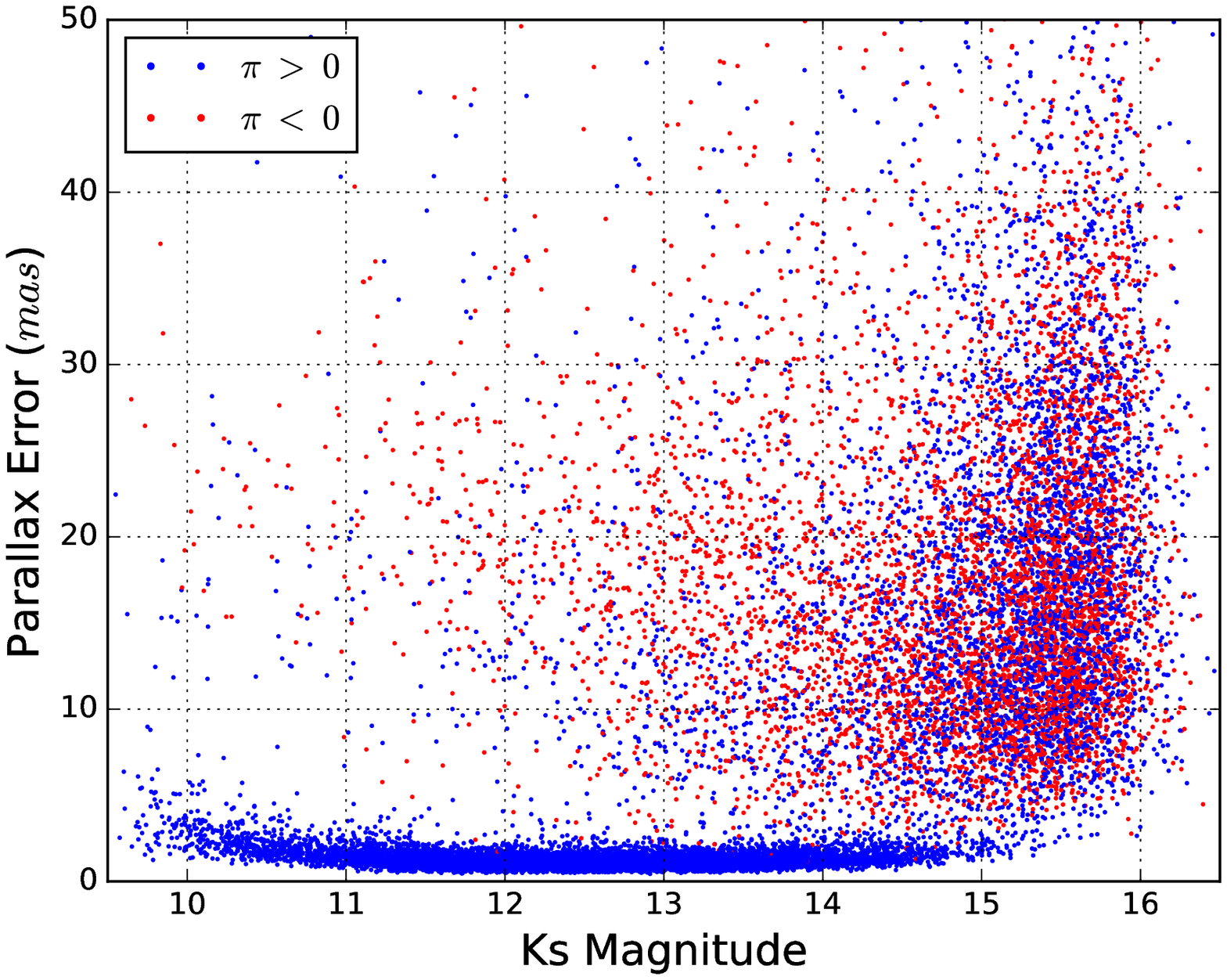,width=1\linewidth,clip=}\\
      \epsfig{file=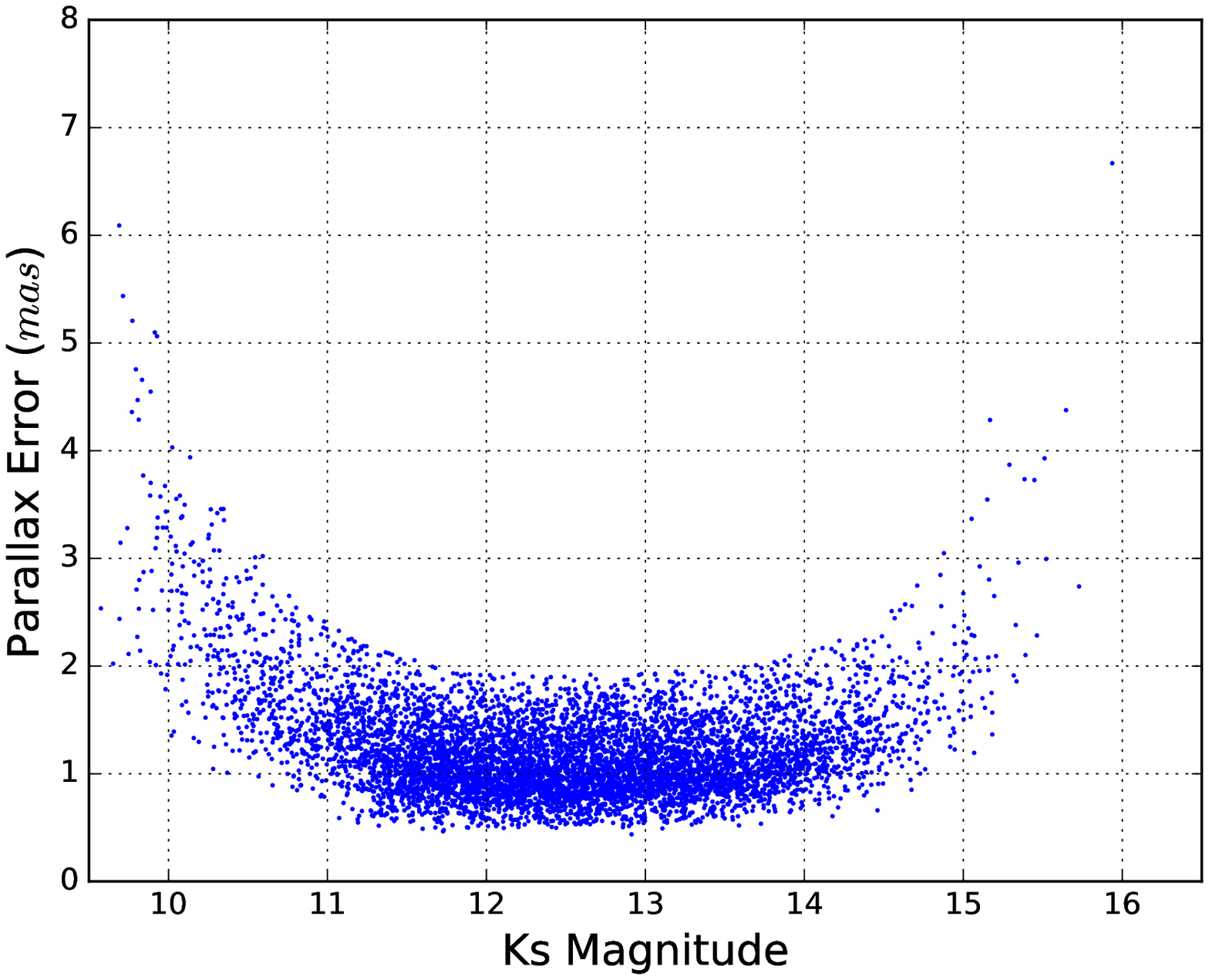,width=1\linewidth,clip=}
    \end{tabular}
    \caption{\textit{upper}: Parallax uncertainty vs. $K_s$ magnitude for sources with detections in two or more pawprints. Only sources with parallaxes measured at greater than $5\sigma$ significance are shown. Blue points are positive parallaxes, red points are negative parallaxes. Since negative parallaxes are not physical, caused by random scatter about true values near zero, we expect the roughly equal number of positive parallaxes in the same region of the plot to be unreliable also. The dearth of negative parallaxes in the lower region of the plot indicates our locus of reliable measurements. \textit{lower}: Our subset of 6935 sources from the upper panel which we deem a reliable selection (see text for details) of parallax measurements.}
    \label{mag_eplx}
  \end{center}
\end{figure}

Parallax uncertainty (averaged between pawprint sets) vs. $K_s$ magnitude for sources with detections in two or more pawprint sets and parallax measured at greater than $5\sigma$ are shown in Figure \ref{mag_eplx} (upper panel). Note the swathe of negative parallaxes (in red on the plot) with large uncertainties and generally faint magnitudes. Since negative parallaxes are clearly not physical but due to random scatter in the measured parallax about true values typically near zero, we expect the roughly equal number of positive parallaxes in the same region of the plot to also be unreliable measurements. The dearth of negative parallaxes in the lower region of the plot indicates our locus of reliable measurements. We attempt to select only reliable parallax measurements as follows.

\begin{figure}
  \begin{center}
    \begin{tabular}{c}
      \epsfig{file=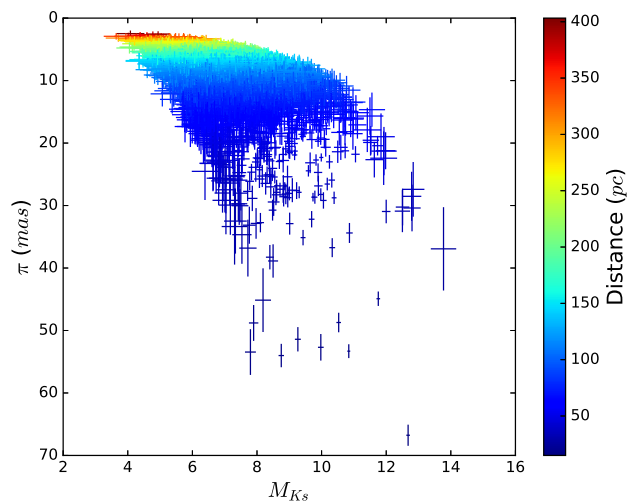,width=1\linewidth,clip=}
    \end{tabular}
    \caption{Absolute $K_s$ magnitude vs. parallax for the 6935 $5\sigma$ parallaxes.}
    \label{plxMKsDist}
  \end{center}
\end{figure}

Testing indicated the following requirements reduced the number of sources in the upper part of the plot significantly:
\begin{itemize}
\item $\ge 5\sigma$ inverse variance weighted average parallax.
\item Parallax measurements from a minimum of two pawprint sets.
\item The two highest weighted measurements agree within $2\sigma$ and 10~mas, are positive, and each measured with at least $2\sigma$ significance.
\item Mean ellipticity (across all epochs) is less than 0.2.
\end{itemize}
To this selection we apply a similar set of cuts to the epm flag =  0 selection above (see Section \ref{epm_flagging}), albeit applied after the averaging of measurements from separate pawprints (including pawprints in adjacent tiles) and across the entire survey at once. Since the parallax dataset is smaller we reduce the width of the magnitude bins to 50 sources. Sources in a magnitude range where the median parallax uncertainty is greater than 5~mas, and those with parallax uncertainty greater than the median for their magnitude plus three times the spread (defined as the larger of either 0.3~mas or the median absolute deviation for their magnitude) are flagged as unreliable. The resultant selection of 6935 "reliable" parallax measurements are shown in Figures \ref{mag_eplx} (lower panel) and \ref{plxMKsDist}. The selection is shown in Table \ref{plxtable}. The distribution of parallax measurements indicate that the bulk of sources are in the $50-200$~pc range.

\subsection{Comparison to TGAS parallaxes}

\begin{figure}
  \begin{center}
    \begin{tabular}{c}
      \epsfig{file=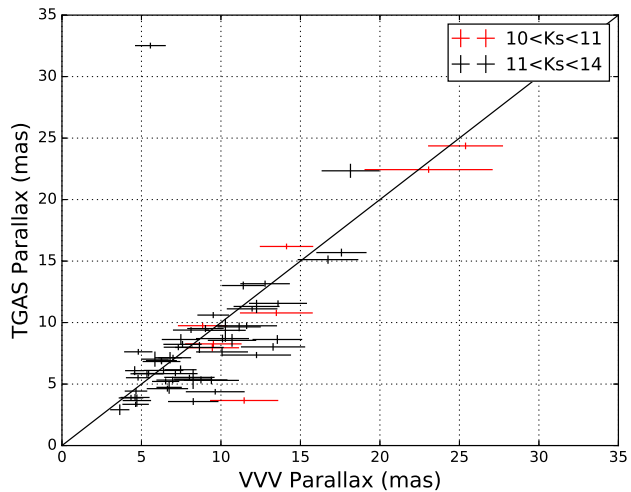,width=1\linewidth,clip=}
    \end{tabular}
    \caption{Comparison of VIRAC and TGAS parallax measurements for 62 common proper motion pairs identified by matching the two catalogues. Black points are those where the VIRAC source is in the $11<Ks<14$ range, red points are those where the VIRAC source is in the $10<Ks<11$ range. The agreement is generally good. We attribute the asymmetry about the equal parallax line to bias introduced by our $5\sigma$ selection, see discussion in text.}
    \label{vvv-tgas_plx}
  \end{center}
\end{figure}

Sources common to TGAS and the VIRAC parallax selection are few. To increase the comparable sample we identify TGAS stars with fainter common proper motion companions in VVV and reason that these should have near identical parallaxes. To identify likely genuine companions we match the two catalogues with a 2\arcmin\ radius and proper motion agreement within 10~mas~yr$^{-1}$ in both $\alpha\cos{\delta}$ and $\delta$. We also require a projected separation below 20,000~au, measured using the TGAS parallax.
We quantify the probability, $p$, that each match is a chance alignment of unrelated objects using a procedure identical to that described in Section \ref{tgas_cpm}. Note that this sample implicitly includes the few sources common to both catalogues. Figure \ref{vvv-tgas_plx} compares TGAS parallaxes with those in VIRAC for the 62 common proper motion matches with $K_s>10$ and $p < 10^{-4}$. The agreement is generally good. The 54 sources with magnitudes $11<K_s<14$ are plotted in black: this magnitude range is typical of the VIRAC catalogue, see Figure \ref{mag_eplx}, and the median parallax uncertainty for these is 1.2~mas, with a small spread. The 8 sources with $10<K_s<11$ are shown in red in Figure \ref{vvv-tgas_plx}. They typically have larger uncertainties due to saturation and they are somewhat over-represented because many binaries have a modest contrast in flux. The extreme outlier (the point at upper left) can be explained as a bogus match, given that physically unrelated proper motion matches with probability $p<10^{-4}$ can occur in a sample of almost 10$^4$ stars.

To quantify the agreement of VIRAC and TGAS parallaxes, we find that 78\% of sources (42/54) agree within 2$\sigma$. This is less than would be expected for Gaussian error distributions with the quoted uncertainties. The 22\% of outlying sources with 2$\sigma$ disagreement are split in the ratio 7\% to 15\% between smaller TGAS distance and larger TGAS distance; there is a similar ratio of 15\% to 39\% for 1$\sigma$ outliers. The excess of outliers and the asymmetry may be understood as a consequence of our $5\sigma$ threshold: in a selection by measurement significance, where the uncertainties are not always precisely measured, we preferentially select sources with small errors such that the uncertainty is on average underestimated. This will cause an excess of outliers relative to the quoted Gaussian error distribution. The asymmetry is simply explained by the fact that sources with true parallax significance below our threshold can scatter in, appearing to the right of the solid line in Figure \ref{vvv-tgas_plx}. Conversely, sources with true parallax significance above our threshold can scatter out, removing them from the left side of the solid line. This is however a complex problem. Other selection effects and biases will also come into play. Note that we did not attempt a parallax measurement for any source with $\mu<20$~mas~yr$^{-1}$, so very distant stars should be removed. Due to the influence of these biases VIRAC parallaxes should be used with caution: their use in combination with colour data is demonstrated below. The most likely distance is often not simply $\pi^{-1}$. \citet{bailerjones15} discuss in detail the problems involved in estimating distances using parallax measurements. The VVVX extension to tfhe VVV survey should allow us to improve the reliability of VIRAC parallaxes over the next few years, even though the formal precision will improve only slightly.

Details of 53 pairs of sources from Figure \ref{vvv-tgas_plx} where the VIRAC component is in the $11<K_{s}<14$ range and excluding the outlier are given in Table \ref{tgas_cpm_plx}.


\section{Discoveries}\label{discoveries}

\subsection{L Dwarfs}\label{ldwf}

One obvious use of the a near-infrared parallax catalogue is the identification of ultracool dwarfs. Even relatively luminous early L-type dwarfs are only visible out to distances of order 100~pc in wide field near-infrared surveys with modest integration times. For this reason, brown dwarfs detected in the VVV survey will often have a measurable parallax, as previously demonstrated by \citet{beamin13} and \citet{smith15}.

\begin{figure}
  \begin{center}
    \begin{tabular}{c}
      \epsfig{file=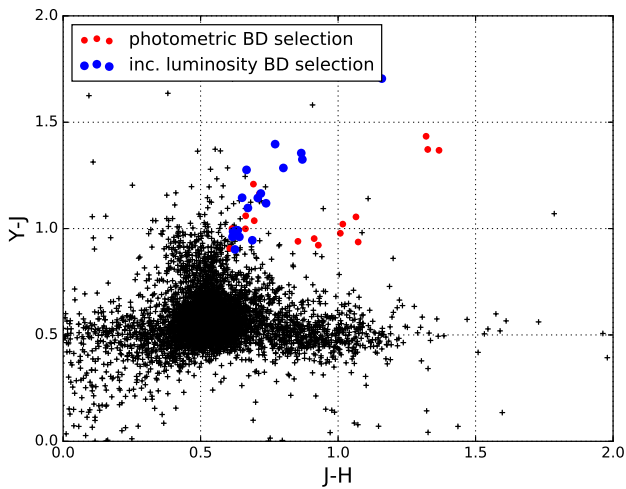,width=1\linewidth,clip=}
    \end{tabular}
    \caption{$J-H$ vs. $Y-J$ for the sources in the VIRAC parallax selection with YJHK detections. Highlighted are the 35 sources which meet our colour selection criteria for L0-T2 dwarfs (circles); those that also meet our luminosity selection criterion are highlighted in blue.}
    \label{Lselect_fig}
  \end{center}
\end{figure}

\begin{figure}
  \begin{center}
    \begin{tabular}{c}
      \epsfig{file=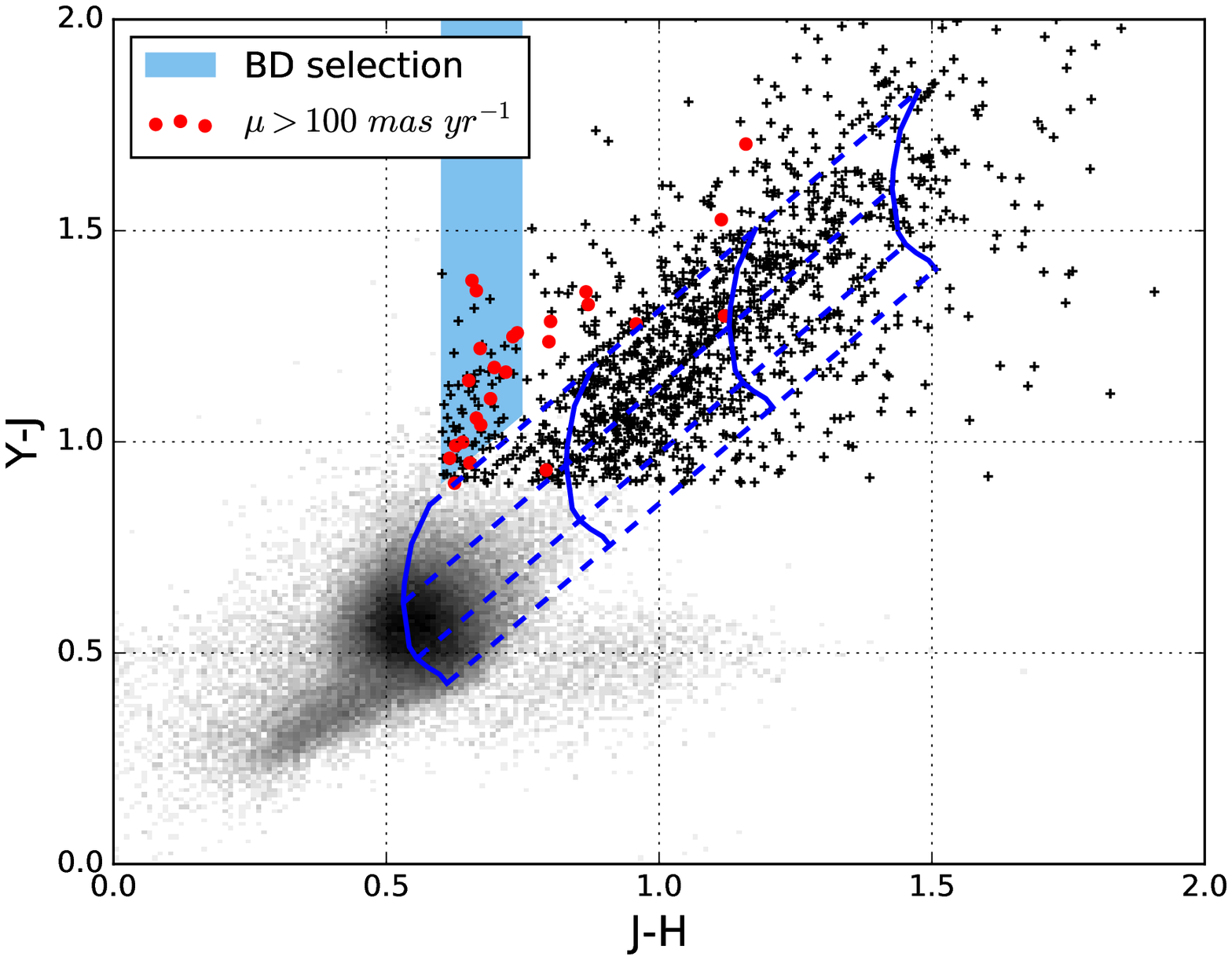,width=1\linewidth,clip=}\\
      \epsfig{file=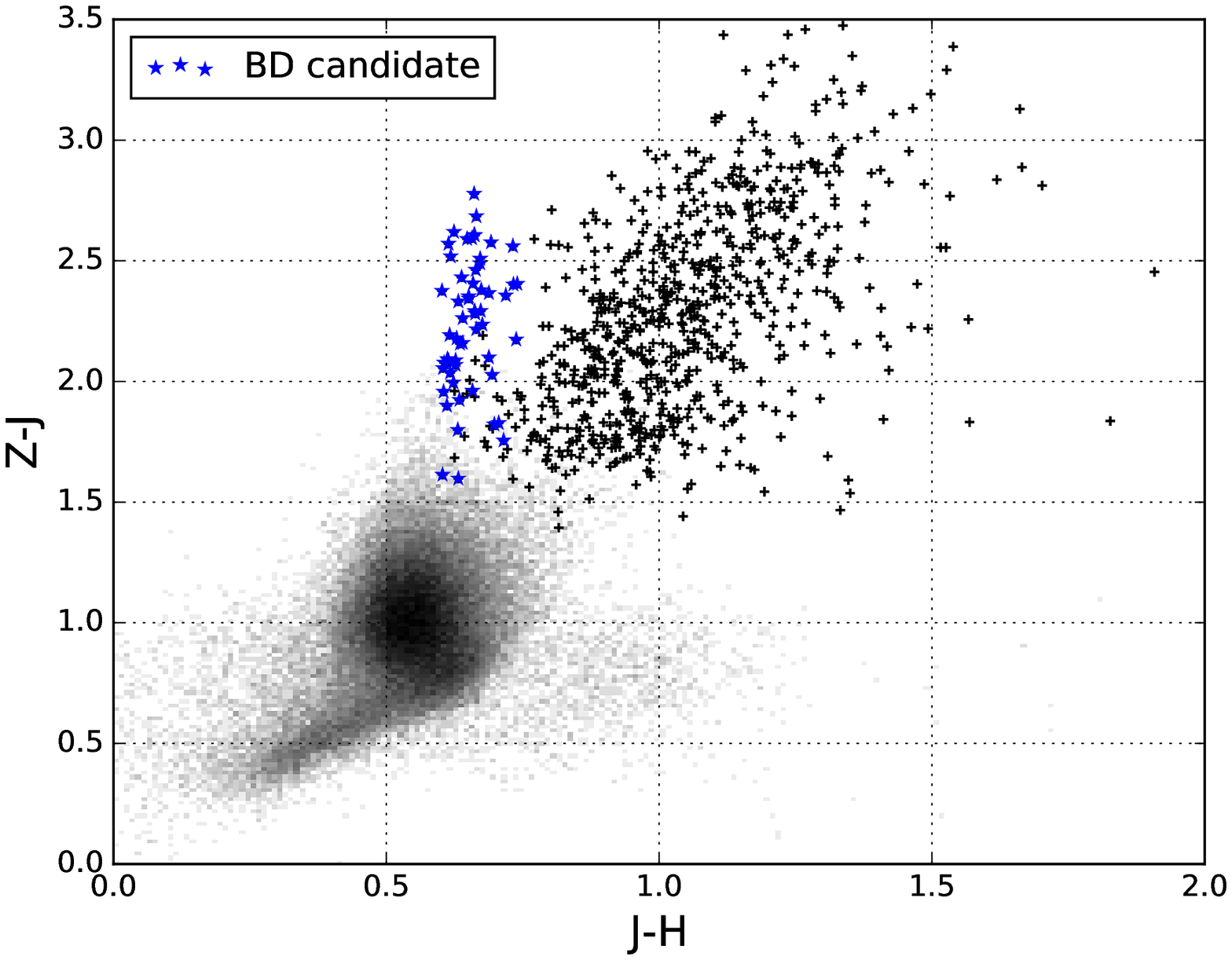,width=1\linewidth,clip=}
    \end{tabular}
    \caption{Density plots of $Y-J$ and $Z-J$ vs. $J-H$ for sources from VIRAC with YJHK detections, $\mu>30$~mas~yr$^{-1}$, $\sigma_{\mu}<$5~mas~yr$^{-1}$ and flagged as 'reliable'. Overplotted in black pluses are those that meet the colour criteria for our brown dwarf selection in the parallax data. \textit{upper}: In red are sources that have $\mu>100$~mas~yr$^{-1}$. Each solid lines shows the M dwarf subtype sequence (M0 bottom, M9 top) for a constant $A_{V}$, and each dashed line shows the reddening sequence for a constant M dwarf subtype at $A_{V}=0$ (left) to $A_{V}=9$ (right). The brown dwarf selection is explained in the text. \textit{lower}: The M dwarfs are further from the early L dwarfs in $Z-J$ but we select on $Y-J$ since many objects are undetected in $Z$. The brown dwarf selection from the upper panel are shown as blue stars.}
    \label{Mseq}
  \end{center}
\end{figure}

We use a set of simple colour selections based on VVV photometry (which are all provided in VIRAC) and the extent of the M dwarf colours tabulated in \citet{Rayala14}:
\begin{description}
\item $Y-J>0.9$, 
\item $J-H>0.6$, 
\item $H-K>0.44$,
\end{description}
and require each of these to be met to be considered a candidate. Further, if a candidate is detected in the $Z$ band we require $Z-J>1.3$ based on the brown dwarf selection of \citet{lodieu07}. These colour selections correspond approximately to brown dwarfs in the subtype range L0-T2, yielding 35 candidate objects (see Figure \ref{Lselect_fig}). The addition of parallax information also allows us to discriminate based on intrinsic luminosity. The \citet{dupuy12} absolute $K_s$ magnitude for an L0 dwarf is approximately 10.4. Allowing for the possibility of equal mass binarity we set our $M_{Ks}$ lower limit at 9.7. The luminosity selection leaves us with 18 promising L0-T2 dwarf candidates, shown in Table \ref{ldwftable} and Figure \ref{Lselect_fig}. Note that since VIRAC parallaxes indicate sources are largely within a few hundred parsecs, we do not take into account the affect of reddening in this selection. A comparison with M dwarf reddening tracks of the upper panel of Figure \ref{Mseq} highlights that our parallax selection is not significantly reddened.

Since the fainter VVV L dwarfs will lack VIRAC parallax measurements (see Figure \ref{mag_eplx}) but often have high proper motions, one might hope to identify additional brown dwarfs in a selection of high proper motion sources. Figure \ref{Mseq} illustrates this approach: the majority of sources with $\mu>30$~mas~yr$^{-1}$ that satisfy the above colour selection are reddened M dwarfs (i.e. relatively distant stars), whereas most of the subset with $\mu>100$~mas~yr$^{-1}$ are (probable) L dwarfs not subject to significant reddening because of their smaller distances. We take the above $Y-J$, $J-H$ and $H-K_s$ brown dwarf colour selection criteria and allow up to $A_{V}=1.5$ of reddening \citep{cardelli89} to define a brown dwarf selection region in the $Y-J$ vs. $J-H$ plane that is limited to a relatively narrow range of $J-H$ colours (see Figure \ref{Mseq}). This will only select early type L dwarfs: later types are more difficult to distinguish from reddened M dwarfs and our intention is to make a reliable rather than complete selection of brown dwarfs. This yielded 66 early L dwarf candidates, see Table \ref{Ltable2}. The separation of these candidates from reddened M dwarfs is clearer in the $Z-J$ vs $J-H$ panel of Figure \ref{Mseq}, for those candidates with a $Z$ detection.

\subsubsection{$\beta$ Cir B}

The young ($\sim 400$~Myr) L1-type age-benchmark companion 
to $\beta$~Cir was discovered in a preliminary version of VIRAC \citep{smith15} by a simple cross-match of L dwarf candidates against HIPPARCOS for stars with similar proper motion and parallax. It is included in Table \ref{ldwftable} with the other L dwarf candidates identified by our colour, parallax and proper motion-based selection described above.

\subsubsection{LTT 7251 B}\label{171004304}

LTT~7251 is a high proper motion G8 type dwarf at $d = 37.6$~pc. It has been extensively studied and boasts a range of accurate chemical abundance measurements courtesy of the HARPS GTO planet search program \citep{adibekyan12} and measurements of a number of other stellar parameters from the Geneva-Copenhagen Survey \citep{casagrande11}. Unfortunately, the current age estimates presented by the Geneva-Copenhagen Survey do not significantly constrain the age of LTT~7251 (approx. 1-10~Gyr). This could potentially be remedied with gyrochronology using a modestly sized telescope to observe this bright star ($V=8.54$, $K_s=6.7$).

\begin{figure}
  \begin{center}
    \begin{tabular}{c}
      \epsfig{file=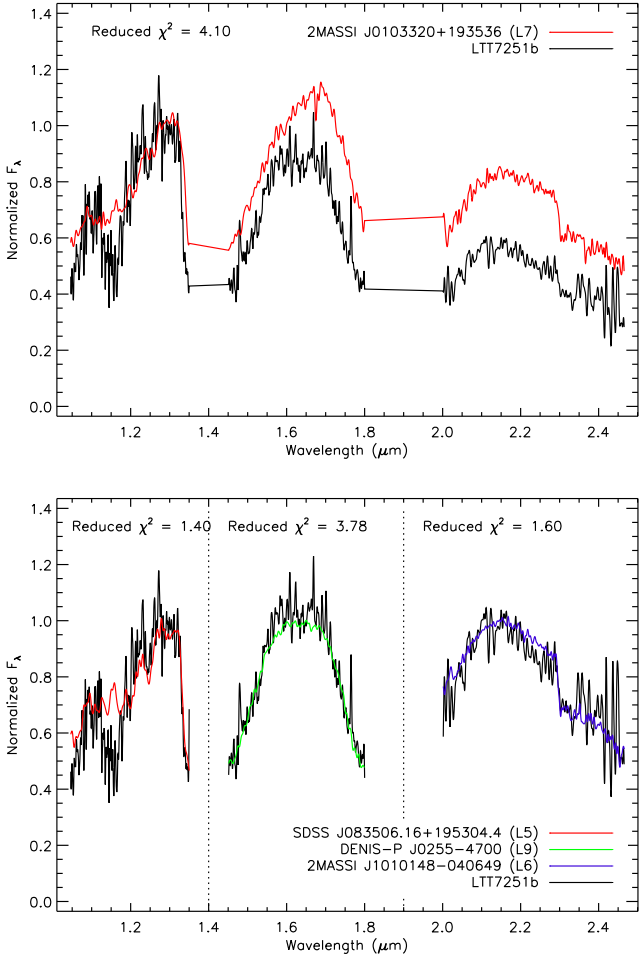,width=1\linewidth,clip=}
    \end{tabular}
    \caption{ARCoIRIS spectrum of LTT 7251 B. \textit{upper}: Plotted against the L7 NIR standard 2MASSI J0103320+193536 observed by \citet{cruz04}. \textit{lower}: Data in the $J$, $H$, and $K$ bands compared with the L5 standard SDSS J083506.16$+$195304.4 \citep{chiu06}, the L9 standard DENIS-P J0255-4700 \citep{burgasser06}, and the L6 standard 2MASSI J1010148-040649 \citep{reid06} respectively. All standard spectra were obtained from the SpeX prism library \citep{burgasser14}.}
    \label{fig:LTT7251B}
  \end{center}
\end{figure}

We have identified a mid-late L dwarf common proper motion companion to LTT 7251 in the VIRAC proper motion data (LTT 7251 B henceforth), located 14.7$\arcsec$ due north of the primary. At the TGAS distance of 37.6~pc, this separation corresponds to 553~au. 
LTT 7251 B was observed on August 20th, 2016 with the Astronomy Research using the Cornell Infra Red Imaging Spectrograph (ARCoIRIS), a cross-dispersed, single-object, longslit, infrared imaging spectrograph, mounted on Blanco 4 m Telescope, CTIO. The spectra cover a simultaneous wavelength range of 0.80 to 2.47 $\mu$m, at a spectral resolution of about 3500 $\lambda$/$\Delta \lambda$, encompassing the entire zYJHK photometric range. The spectrum was taken with 16 exposures, each 180 sec in ABBA dithering pattern for sky background subtraction.  The average airmass was 1.4 and seeing 1.2\arcsec. We observed the telluric A0 V standard HD163336 immediately after the target. The data were reduced using the Spextool IDL package (version 4.1, \citealt{cushing04}), a new suite of data reduction algorithms specifically designed for the data format and characteristics of ARCoIRIS by Dr. Katelyn Allers \footnote{\href{http://www.ctio.noao.edu/noao/node/9701}{www.ctio.noao.edu/noao/node/9701}}. Telluric correction and flux calibration of the post-extraction spectra are achieved through the xtellcorr IDL package \citep{vacca03}. The reduced spectrum is shown in Figure \ref{fig:LTT7251B}.

This source did not fulfil the L dwarf selection criteria described in the previous section because it lacks $Z$ and $Y$ photometry in the standard VVV catalogue products of the CASU pipeline, owing to proximity to a diffraction spike associated with the much brighter primary and the slightly poorer spatial resolution of VISTA in the $Z$ and $Y$ passbands. The colour $J-H=0.57$ also failed our selection criterion, though this may simply be because of the significant uncertainty (0.14 mag). We were able to extract $Z$ and $Y$ photometry from the images by using small photometric apertures (1.414$\arcsec$ diameter for $Y$, 1$\arcsec$ diameter for $Z$) and carefully placing a sky background aperture. These data and VVV DR4 source magnitudes in a 1\arcsec~ diameter aperture are given in Table \ref{LTT7251_table}. The source also lacks a useful parallax measurement because it is relatively faint $K_s=15.47$). However, there are no other {\it reliable} high proper motion sources in the same 1.6~deg$^2$ VVV tile with proper motion vectors consistent with the primary within a 15~mas~yr$^{-1}$ tolerance, so the probability of a chance projection of a high common proper motion companion within 15$\arcsec$ is below $3.4\times 10^{-5}$.

\begin{table}
\begin{center}
\caption{Details of the LTT 7251 system.}
\label{LTT7251_table}
\begin{tabular}{|l|c|c|l|}
 & A & B \\
\hline
$\alpha$ & 18h15m49.14s & 18h15m49.14s & 2015.0 \\
$\delta$ & -23d48m59.92s & -23d48m45.23s & 2015.0 \\
$\mu_{\alpha\cos\delta}$ & $64.83\pm0.08$ & $76.3\pm6.9$ & mas~yr$^{-1}$ \\
$\mu_{\delta}$ & $-166.27\pm0.05$ & $-172.0\pm5.1$ & mas~yr$^{-1}$ \\
$\pi$ & $26.59\pm0.31$ &  & mas \\
$B$  & $9.29\pm0.02$ & & mag \\ 
$V$  & $8.54\pm0.01$ & & mag \\ 
$Z$  & & $19.85\pm0.20$ & mag \\ 
$Y$  & & $18.10\pm0.10$ & mag \\ 
$J$  & $7.18\pm0.02$ & $16.93\pm0.09$ & mag \\ 
$H$  & $6.80\pm0.02$ & $16.36\pm0.11$ & mag \\ 
$K_s$ & $6.73\pm0.03$ & $15.47\pm0.05$ & mag \\ 
$\rho$ & \multicolumn{2}{|c|}{14.7} & arcsec \\
$\rho$~proj. & \multicolumn{2}{|c|}{553} & au \\
\hline
\end{tabular}
\end{center}
Note: LTT 7251 A astrometry is that of TGAS, optical photometry is from the Tycho 2 catalogue \citep{hog00}, NIR photometry is from 2MASS. The LTT 7251 B Z and Y band photometry were measured from the reduced VVV images (see text), the J and H band photometry are VVV DR4 aperMag1, the K$_s$ band photometry is the average aperMag2 across all epochs used by our astrometric pipeline. LTT 7251 B astrometry is from VIRAC and is relative to the astrometric reference stars but the A component astrometry is absolute hence the moderate difference in proper motion.
\vspace{0.5cm}
\end{table}

A fit across the whole NIR spectrum gives an L4 spectral type (best fit template 2MASS J21580457-1550098, $\chi^2_{red}= 3.197$). However, the L4 template doesn't really reproduce the morphology of the $J$ and $K$ bands and the absolute magnitudes, $M_{J}=14.05$, $M_{H}=13.48$, $M_{Ks}=12.6$, imply a later type. The L7-L9 templates give a much better fit to the shape of the spectrum, but the overall 1--2.5~$\mu$m slope is bluer (hence a larger $\chi^2$, see Figure \ref{fig:LTT7251B}).\\
If we fit the $J$, $H$, and $K$ bands separately we obtain later spectral types:
\begin{description}
\item J-band best fit: L5 SDSS J083506.16$+$195304.4 ($\chi^2_{red}=1.3$)
\item H-band best fit: L9 DENIS-P J0255-4700 ($\chi^2_{red}=3.2$)
\item K-band best fit: L6 2MASSI J1010148-040649 ($\chi^2_{red}=1.6$)
\end{description}
The average spectral type when fitting $J$, $H$, and $K$ separately is L7, which agrees with the visual matching to the L7 spectral standard. It's mildly low metallicity is likely the explanation for the slightly blue colour of LTT 7251 B and the slight under-luminosity in the $K_s$ band ($\sim0.2$~mag) relative to a field L7 dwarf.

The list of accurate chemical abundances available for LTT 7251 A\/B and potential for a reasonably accurate age determination via gyrochronology make the system a promising benchmark, a test of brown dwarf atmospheric forward grid models and retrieval methods (see e.g. \citealt{burningham17}).

\subsubsection{VVV J12564163-6202039}\label{357154962}

VVV J12564163-6202039 (VVV~1256-6202 hereafter) is the only genuine source we identified through visual inspection of sources with $\mu>1\arcsec~yr^{-1}$, and $10<\sigma_{\mu}$ (mas~yr$^{-1})<30$, see section \ref{addvisinspect}. It is listed in the 2MASS Point Source catalogue as 2MASS J12564352-6202041, detected in $J$ only. VVV~1256-6202 is severely blended with a background source in the 2010 epochs; this is likely the cause of the relatively high uncertainty on the proper motion (hence epm flag=2 for both measurements). The pipeline proper motions for the source are $-1112\pm12$ and $-13\pm16$~mas~yr$^{-1}$ in $\alpha\cos\delta$ and $\delta$ respectively. The pipeline median $K_s$ magnitude is $15.77\pm0.04$. Note that while we use robust methods where possible, pipeline outputs for this source are swayed to some extent by the blended early epochs. This source does not have a parallax in VIRAC; measurements were attempted in the two overlapping pawprint sets separately but they did not agree.
To try to improve on the pipeline values with a more bespoke solution, we use VVV $K_s$ band observations of the source from all pawprint sets for a single fit, omitting data from the 2010 observing season. The median $K_s$ band magnitude for these 82 epochs is $15.737\pm0.056$. Using astrometric reference sources this time drawn from within 1\arcmin\ radially about the target but an otherwise similar fitting procedure we measure proper motions of $-1116.3\pm4.1$ and $3.9\pm4.0$~mas~yr$^{-1}$ in $\alpha\cos\delta$ and $\delta$ respectively and a parallax of $5.3\pm7.1$~mas.

The $ZYJH$ photometry that we have included in VIRAC taken from VVV DR4 at the VSA, mostly corresponding to the first set of multi-filter data taken at the beginning of the survey. This source is blended with the background source in the early observations and hence shows null (99.999) $ZYJH$ detections in VIRAC. We therefore obtained the $ZYJHK_s$ tile catalogue photometry produced from the second set of multi-filter observations made at the end of the survey. These magnitudes are given in Table \ref{357154962_table}. This field was also observed as part of the VPHAS$+$ survey \citep{drew14}, giving us a $i$ bandpass psf fit magnitude from DR2\footnote{\href{http://cdsarc.u-strasbg.fr/viz-bin/Cat?II/341}{cdsarc.u-strasbg.fr/viz-bin/Cat?II/341}} (also given in Table \ref{357154962_table}, \citealt{drew16}) but non-detections in the $r$ and $H\alpha$ observations. The source was detected in the $J$ band in 2MASS ($16.1\pm0.1$, photometric quality flag B), but not in the $H$ and $K_s$ bands (the upper limits on these magnitudes are $15.5$ and $15.3$ respectively).

\begin{table}
\centering
\caption{Details of VVV~1256-6202. All fluxes are on the Vega system.}
\label{357154962_table}
\begin{tabular}{|l|c|l|}
\hline
$\alpha$ & 12h56m41.63s & 2012.0 \\
$\delta$ & -62d02m03.93s & 2012.0 \\
$\mu_{\alpha\cos\delta}$ & $-1116.3\pm4.1$ & mas~yr$^{-1}$ \\
$\mu_{\delta}$ & $3.9\pm4.0$ & mas~yr$^{-1}$ \\
$i$  & $19.70\pm0.08$ & mag \\
$Z$  & $17.86\pm0.03$ & mag \\
$Y$  & $17.01\pm0.02$ & mag \\
$J$  & $16.10\pm0.01$ & mag \\
$H$  & $15.89\pm0.02$ & mag \\
$K_s$ & $15.72\pm0.03$ & mag \\
\hline
\end{tabular}
\end{table}

\begin{figure}
  \begin{center}
    \begin{tabular}{c}
      \epsfig{file=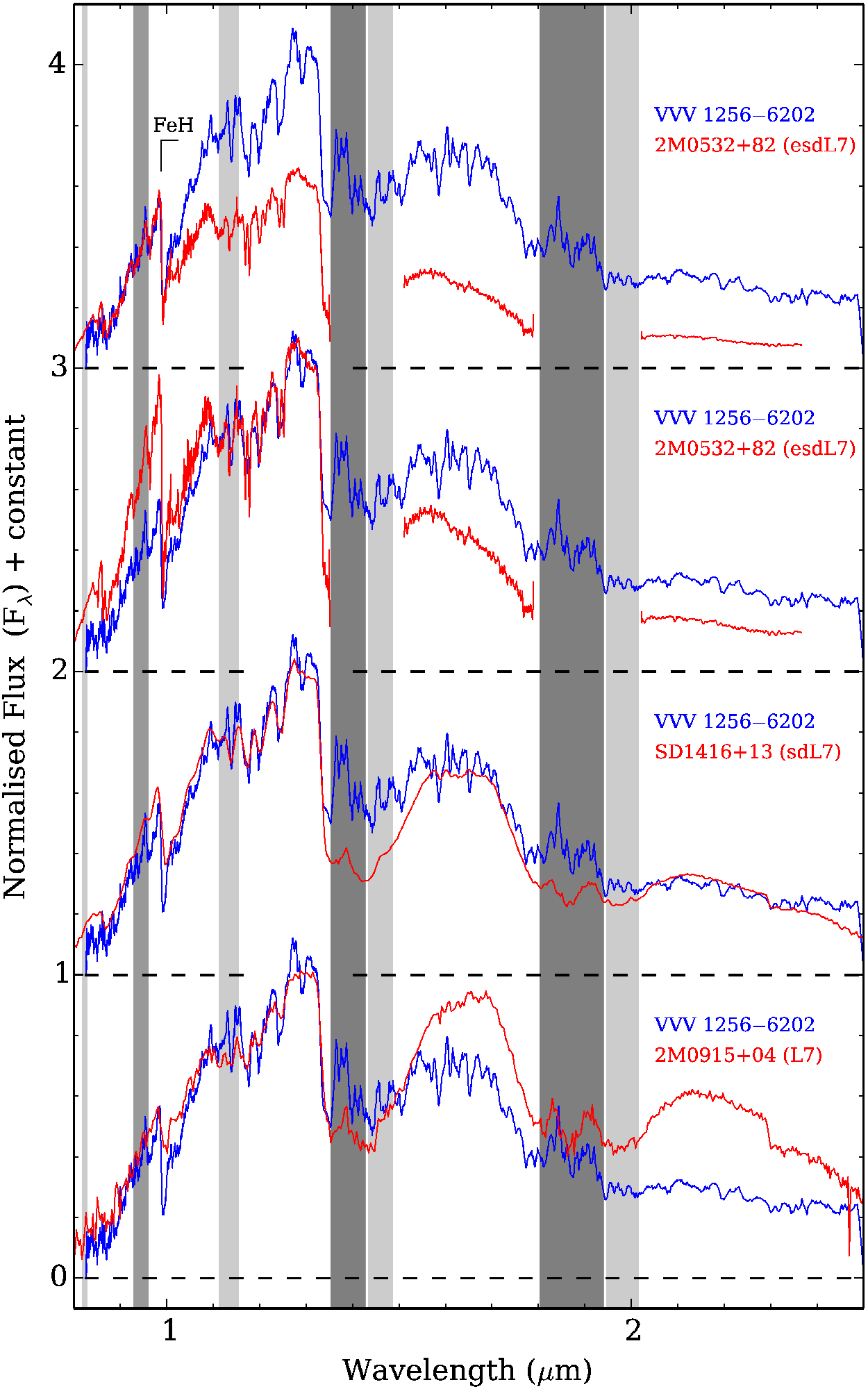,width=1\linewidth,clip=}
    \end{tabular}
    \caption{FIRE spectrum of VVV 1256-6202 compared to known L dwarfs and subdwarfs. The spectrum of VVV 1256-6202 is smoothed by 5 pixels. Telluric absorption regions are highlighted in grey (darker grey represents regions of stronger telluric features) and have been corrected for all spectra.
The upper two panels show comparison with an esdL7 dwarf, normalised in the far red optical (top panel) and the $J$ bandpass (2nd panel) respectively to illustrate the strength of FeH absorption. The spectrum of 2MASS J09153413+0422045 (2M0915+04) is from \citet{burgasser07}; SDSS J141624.08+134826.7 (SD1416+13) is from \citet{schmidt10}; and 2MASS J05325346+8246465 (2M0532+82) is from \citet{burgasser03}. The spectrum of 2M0532+82 at 1.008-1.153 $\mu$m wavelength is missing. The best-fitting BT-Settl model spectrum of 2M0532+82  ($T_{\rm eff}$ = 1600 K, [Fe/H] = -1.6, and log $g$ = 5.25) is plotted to fill the gap.
} 
    \label{fig:2M1256-62}
  \end{center}
\end{figure}

VVV~1256-6202 is a significant outlier from the locus of normal dwarf stars in $K_s$ band reduced proper motion (see Figure \ref{hpm_rpm}). It has colours consistent with those of known extreme L subdwarfs (esdL types) (\citealt{kirkpatrick14}; \citealt{zhang17a}; \citealt{burgasser04}). On figure 2 of \citet{zhang17a} VVV~1256-62 would fall in the $T_{eff}=2300-2400~K$ range, of the $[Fe/H]\approx{}-1.0$ Bt Settl grids \citep{allard14}, after allowing for the positive Vega to AB magnitude offset in $i$ of $\sim$0.37~mag \citep{hewett06}. This is promising evidence to suggest an esdL type.

To test this, we obtained a 0.8--2.5~$\mu$m spectrum (see Figure \ref{fig:2M1256-62}) with the Folded-port InfraRed Echellette (FIRE) spectrograph \citep{simcoe08} mounted on the Magellan Baade Telescope at Las Campanas Observatory, Chile.
The spectrum was taken in the low resolution prism mode with a 1$\arcsec$ slit width, yielding spectral resolution R$\approx$250, in conditions of mediocre seeing (1.2 to 1.5$\arcsec$ at full-width-half-maximum) so the data are not of the best quality. 
The total exposure time was 20~min, nodding along the slit with individual exposures of 60~s duration. The star HIP~58411 was used for telluric correction and the data were reduced with the {\sc FIREHOSE} software package using standard methods. Wavelength calibration was done with a NeAr lamp shining on a screen in the telescope dome, external to the instrument.

In Figure \ref{fig:2M1256-62} we show a comparison of the spectrum with those of a normal L dwarf (dL7) and two subdwarfs (sdL7 and esdL7 types) and a BT-Settl model with $T_{\rm eff}$ = 1650 K, [Fe/H] = -1 and log $g$ = 5.25 \citep{allard14}. Using visual inspection, the sdL7 type provides the best match at $\lambda > 1$~$\mu$m, though the esdL7 fits slightly better at shorter wavelengths. Our adopted type is sdL7, though we note that several features suggest it is toward the low metallicity end of the sub-dwarf range:
stronger FeH absorption than the sdL7 reference sources at 0.99~$\mu$m, less prominent flux peaks in the $H$ and $K$ bandpasses and the blue broadband $J-K$ colour more consistent with esdL type than sdL type. The range of metallicities for sdL types
is $-1.0<[Fe/H]<-0.3$, whereas for the esdL7 source
2MASS J0532+8246 it is $[Fe/H] = -1.6$ \citep{zhang17a}, so the evidence suggests a metallicity close to -1 dex.

The absolute magnitudes of an sdL7-type ultracool dwarf are
$M_J=13.14 \pm 0.40$ and $M_H = 12.92 \pm 0.40$ \citep{zhang17a}. This implies a distance of $39 \pm 8$ ~pc for VVV~1256-6202. This object could have a distance of $55 \pm 11$ pc, if the object is an unresolved equal mass binary. The parallax of $5.3 \pm 7.1$~mas implies $d>51$~pc at 2$\sigma$ confidence or $d>38$~pc at 3$\sigma$ confidence.

Given that this source is substantially fainter than almost all sources with a 5$\sigma$ VIRAC parallax measurement we do not give the distance discrepancy much import. Nonetheless, the $i-J$ colour of VVV~1256-6202 is much bluer than the sdL7 dwarf SDSS~1416+13, implying $T_{eff} \approx$~2300-2400~K as noted earlier, much warmer than the 1650~K fit to the BT-Settl model.
Inspection of the VPHAS+ image and catalogue for the field indicates that the $i$ flux measurement is reliable, by comparison with adjacent sources. An alternative interpretation more consistent with the broadband colours and the limits on parallax is that the spectral type is esdL1. The spectrum matches an esdL1 type well at $\lambda > 1.4~\mu$m but there is a very poor fit at shorter wavelengths (not shown). We therefore adopt the sdL7 type but we note that a better quality spectrum is desirable, both to check this and to enable more detailed analysis. 

The tangential velocity implied by the sdL7 spectrophotometric distance is $v_{tan} \approx 205$~km/s. Lacking a radial velocity datum, we considered a wide range of possible values and calculated the Galactic space velocity components ($U,V,W$) for each. The equations of \citet{bensby03} then showed that
this sub-dwarf can plausibly be a member of the thick disc or the halo. Thin disc membership has negligible probability. An esdL1 type would imply greater distance, higher velocity and hence halo membership. However, we note that that proper motion of this source is almost exactly parallel to the Galactic plane such that the $W$ velocity component is very small for any plausible distance. This tends to favour thick disc membership and the sdL7 interpretation.

With metallicity near [Fe/H]=-1 and spectral type sdL7,
VVV~1256-6202 is only the fourth mid-late L-type subdwarf or extreme subdwarf to be discovered. It shares with ULAS~1338-0229 \citep{zhang17a} the title of the latest spectral type near the 
[Fe/H]=-1 mark. Comparison with figure 9 of \citet{zhang17b} indicates that this ultracool dwarf has a mass below $M\approx 0.079~M_{\odot}$, and below the (metallicity-dependent) hydrogen-burning minimum mass threshold, assuming a temperature of 1650~K, the same as the ULAS J1338$-$0229 \citep{zhang17b}.
The relative brightness and proximity of this subdwarf make it a good target for future investigation.

\subsection{New Members of the 25~pc Sample}\label{solneighbours}

\begin{figure}
  \begin{center}
    \begin{tabular}{c}
      \epsfig{file=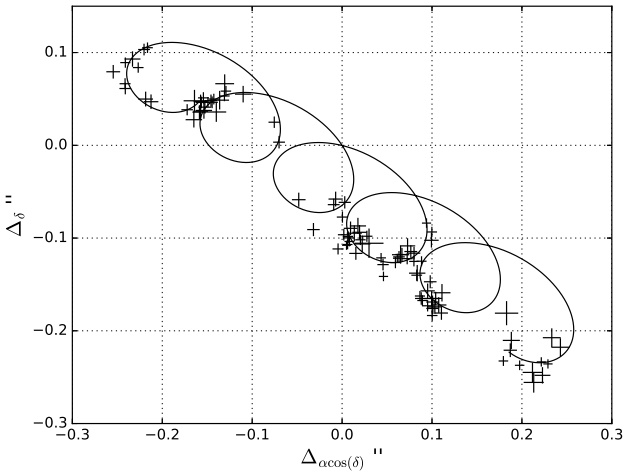,width=1\linewidth,clip=}
    \end{tabular}
    \caption{The astrometric fit of VVV~J14115932-59204570, a white dwarf candidate (source \#2 in Table \ref{dlt25table}) and the nearest VIRAC source at $d=15$~pc.} 
    \label{392515914_pmplx}
  \end{center}
\end{figure}

We find ten objects whose VIRAC parallaxes make them members of the 25pc volume limit sample; these are shown in Table \ref{dlt25table}. Of the ten sources, only three are known high proper motion objects: \#5 is VVV BD001 \citep{beamin13}; \#8 is 2MASS J18015266-2706540, original identification by \citet{lepine08}; and \#10 is 2MASS J18084755-2632395, originally identified by \citet{sumi04}. Of those three only one, VVV BD001, was a known member of the 25pc volume limited sample. In Figure \ref{392515914_pmplx} we show the astrometric fit for source \#2, VVV~J14115932-59204570, the object with the largest VIRAC parallax. VVV~J14115932-59204570 is relatively faint and blue; its location in the reduced proper motion diagram of Figure \ref{hpm_rpm} suggests it is a white dwarf. Based on its near infrared colours and luminosity and comparison with \citet{hewett06} (table 13 in particular), VVV~J14115932-59204570 is most consistent with a $T_{eff}\approx6500-7000$K pure-H white dwarf.

\section{Pan-Galactic applications}\label{pan-galactic}
\subsection{Galactic Rotation Curve}\label{galrotcurve}

To demonstrate the usefulness of VVV proper motions out to large distances and across large areas we attempt to measure the galactic rotation curve tangential to the line of sight in the $l=300^{\circ}$ direction using reasonably well calibrated red clump giant branch standard candles (\citealt{lopez02}, \citealt{hawkins17}). This is merely a demonstration of potential since a true measurement will require a precise relative to absolute correction of the proper motions using the Gaia absolute reference frame, following the forthcoming Gaia 2nd Data Release.

We use VIRAC sources flagged as reliable at $299.5<l<300.5$ and $0.5<|b|<1.0$ with $10<J<20$ and $10<K_s<20$. On this sample we perform an approximate relative to absolute proper motion correction by using the median proper motion of a Besan\c{c}on synthetic stellar population \citep{robin03} in the direction of our VVV sample and at a $K_s$ magnitude range identical to that of our astrometric reference source selection ($12.5<K_s<16.0$). We use a diffuse extinction parameter of 2.0~mag~kpc$^{-1}$. Our relative to absolute corrections in $\mu_{l}$ at $l=300^{\circ}$ are -6.0~mas~yr$^{-1}$ and -5.9~mas~yr$^{-1}$ for $b=0.75$ and $b=-0.75$ respectively.

\begin{figure}
  \begin{center}
    \begin{tabular}{c}
      \epsfig{file=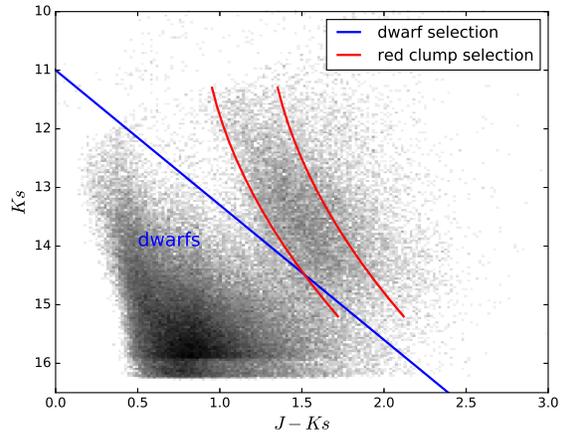,width=1\linewidth,clip=}
    \end{tabular}
    \caption{A log density plot of our $l=300^{\circ}$ sample in $J-K_s$ vs. $K_s$ colour magnitude space. Overplotted are our simple linear dwarf identification cutoff, and the $\pm0.2~mag$ bounds around our fitted red clump giant tract that we use to select red clump giant stars. The bimodal feature at K$_s\approx$16 is due to different magnitude cutoffs for the reliability flag in the different fields.}
    \label{rcg_colour-mag}
  \end{center}
\end{figure}

We perform an initial approximate dwarf rejection for this Galactic coordinate range of $K_s>2.3(J-K_s)+11$ (see Figure \ref{rcg_colour-mag}). The remaining initial giant candidates in the range $11.3<K_s<15.5$ we split into 0.3~mag wide bins and find the approximate location of the peak of the $J-K_s$ distribution for each bin. Using a second order polynomial we fit the $J-K_s$ peak location to the median $K_s$ magnitude for each bin to define our red clump giant tract. Sources with $J-K_s$ within 0.2~mag of this tract are our red clump giant selection (see Figure \ref{rcg_colour-mag}).

\begin{figure}
  \begin{center}
    \begin{tabular}{c}
      \epsfig{file=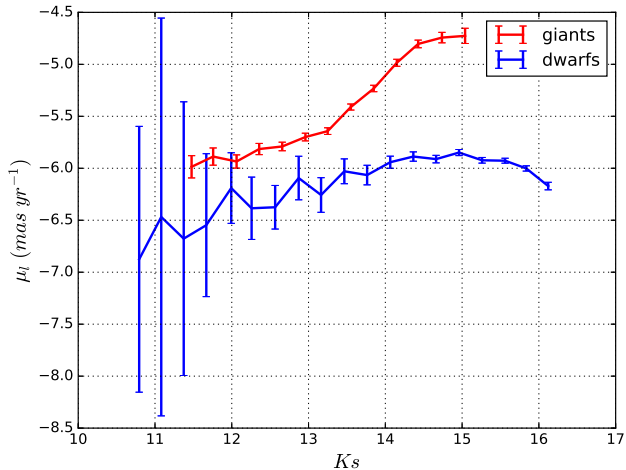,width=1\linewidth,clip=}
    \end{tabular}
    \caption{Binned $K_s$ magnitudes and their median $\mu_{l}$ proper motion values for the dwarf and red clump giant selections shown in Figure \ref{rcg_colour-mag}. }
    \label{rcg-dwf_mag-pml}
  \end{center}
\end{figure}

With our selected red clump giant branch we estimate distances using equations 8 and 9 of \citet{lopez02}, and the updated $M_{K_s}$ and intrinsic $J-K_s$ colours of the red clump giants from \citet{hawkins17}. We then measure the median $\mu_{l}$ and distance in each $0.3$~mag wide $K_s$ magnitude bin. Figure \ref{rcg-dwf_mag-pml} shows the median $\mu_{l}$ vs. $K_s$ magnitude and compares these to the equivalent for our earlier dwarf selection to show that we do indeed measure distinct proper motion distributions for each population. Note that the average motion for dwarfs in the $14<K_s<16$ range is approximately equal to the relative to absolute correction applied earlier, consistent with these comprising the bulk of our astrometric reference sources.

\begin{figure}
  \begin{center}
    \begin{tabular}{c}
      \epsfig{file=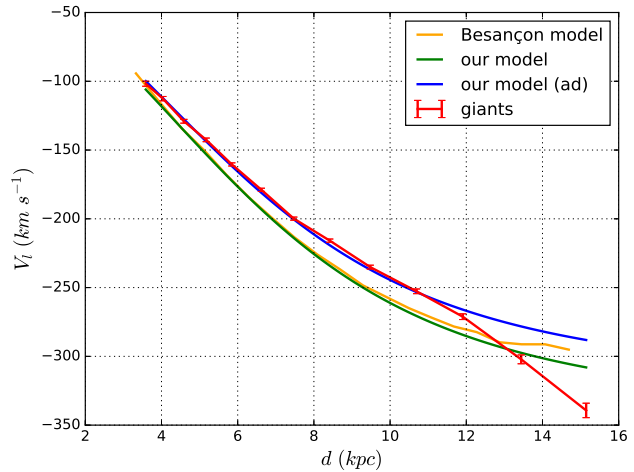,width=1\linewidth,clip=}
    \end{tabular}
    \caption{Tangential velocity in the $l$ direction versus distance for our red clump giant selection, and the Besan\c{c}on synthetic population and simple constant rotation model at the same $l=300^{\circ}$. The discrepancy at $r>12~kpc$ is likely to be due to a bias in selecting only the brightest red clump giants at the faint end of our sample.}
    \label{rcg_dist-vl}
  \end{center}
\end{figure}

We then convert those median $\mu_{l}$ values to a median tangential velocity in $l$ using the median distances for each bin and show the resulting velocity vs. distance curve in Figure \ref{rcg_dist-vl}. Also shown are the equivalent curves for: (i) red clump giants in the Besan\c{c}on model acquired previously; (ii) a simple model which assumes a flat rotation curve with $V_{0}=220~$km~s$^{-1}$; (iii) a flat rotation curve with $V_{0}=205~$km~s$^{-1}$, motivated by the fact that asymmetric drift in the solar neighborhood reduces the orbital velocity of mature stars in the thin disc by an average of 15~km~s$^{-1}$ (e.g. \citealt{bensby03}), though the magnitude of the effect is expected to vary with Galactocentric radius and stellar age (e.g. \citealt{robin03}). Both flat rotation models adopt an $8.5$~kpc galactocentric distance and a solar motion of $(U, V, W)_{solar}=(11.1, 12.24, 7.25)$~km~s$^{-1}$ \citep{schonrich10} relative to the local standard of rest. Our velocity curve agrees fairly well with the models, the best model being the flat rotation curve with asymmetric drift. Alternatively, the data would agree very well with the other two models if the relative to absolute correction in $\mu_{l}$ were changed by only 0.5~mas~yr$^{-1}$, which is very possible given that this quantity is somewhat uncertain. The uncertainty estimates on the velocities are encouragingly small\footnote{For the purpose of this demonstration we neglect to include the contribution of the distance uncertainty in the tangential velocity uncertainty.} (approx. 1.5~km~s$^{-1}$ for $3.5<d~<10.0$~kpc, and still below 5~km~s$^{-1}$ at approx. 14~kpc). We note that our Besan\c{c}on model-based relative to absolute correction is dominated by relatively nearby dwarf stars with $K_s \approx 15$--16 (see figure \ref{rcg_colour-mag}). While there is some circularity in looking at the Besan\c{c}on prediction for distant giants, the results are still sensitive to changes in the Galactic rotation curve along the line of sight. The turn down in our data which starts at around $12~kpc$ ($K_s \approx 14.5$) is due to the decline in the numbers red clump giants at fainter magnitudes and increasing contamination by dwarf stars. A push to slightly fainter magnitudes by using point spread function fitted astrometry/photometry would improve this.

\subsection{Relative to Absolute Correction with Galaxies}\label{rel2abs}

That our Galactic rotation curve agrees reasonably well with the models suggests that our relative to absolute proper motion correction at $l=300^{\circ}$ based on average motions of the Besan\c{c}on synthetic population was accurate. To investigate how well this approach might work at other Galactic locations, and to demonstrate the usefulness of VIRAC for investigating the space motions of populations inside the Galactic bulge, we compare proper motions of a Besan\c{c}on synthetic model at the same magnitude range as the VIRAC astrometric reference sources ($12.5<K_s<16.0$) to those of VIRAC in tile b201 ($l=350.8^{\circ}$; $b=-9.7^{\circ}$), one of the least dense fields in the VVV survey. Unlike in Section \ref{galrotcurve} (where we were sampling the inner disk) we are able to identify a population of external galaxies in tile b201 due to the relatively low stellar density (see Figure \ref{vvv_source_density}). We expect to still be able to select useful numbers of external galaxies at approximately $b=-3.5^{\circ}$, less so closer to the equator. External galaxies should have negligible measurable absolute proper motion, and hence their average relative proper motion in VIRAC tells us the correction that must be applied to the relative motions of nearby (on the sky) objects to place them on an absolute frame (see e.g. \citealt{smith14}).

\begin{figure}
  \begin{center}
    \begin{tabular}{c}
      \epsfig{file=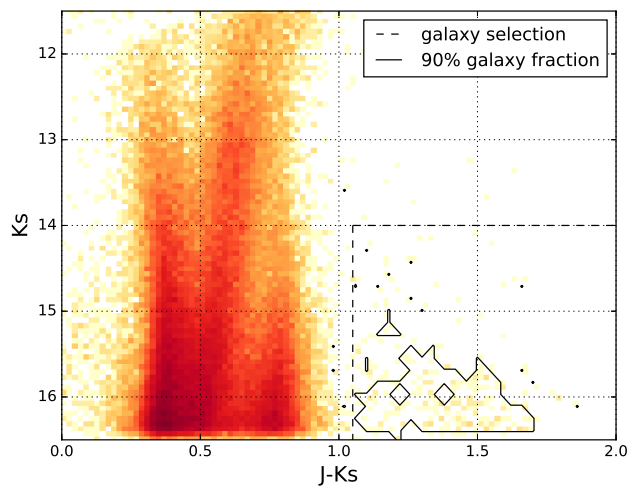,width=1\linewidth,clip=}
    \end{tabular}
    \caption{$J-K_s$ colour vs. $K_s$ magnitude for sources flagged as reliable from tile b201, also shown is our colour-magnitude selection which defines our external galaxy selection for this tile. The contour shows the region inside which galaxy classifications accounted for 90\% of sources that were classified in all bands as either stellar or galaxies.}
    \label{jmk-k_gal}
  \end{center}
\end{figure}

Figure \ref{jmk-k_gal} shows $J-K_s$ colour vs. $K_s$ magnitude for VIRAC sources flagged as reliable in tile b201 and the region from which we select probable external galaxies is indicated. The contour shows the region in which galaxy classifications accounted for 90\% of sources that had a consistent "star" or "galaxy" morphological classification in all bands. This selection yields 797 probable external galaxies.

\begin{figure}
  \begin{center}
    \begin{tabular}{c}
      \epsfig{file=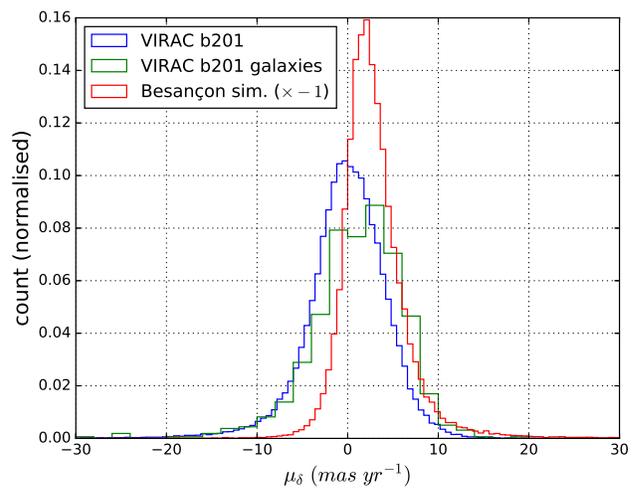,width=1\linewidth,clip=}
    \end{tabular}
    \caption{The distributions of VIRAC $\mu_{\delta}$ for VVV tile b201 as a whole and a selection of external galaxies in tile b201 (see text). Also shown is the additive inverse of the absolute $\mu_{\delta}$ of a Besan\c{c}on synthetic stellar population at the same Galactic location. As expected, the average VIRAC relative motion of tile b201 as a whole is zero. The average relative motion of the selection of external galaxies, which should have effectively zero absolute motion, is $4.8\pm1.0$~mas~yr$^{-1}$. The additive inverse of the average absolute motion of the Besan\c{c}on stellar model should equal this value but it differs by $\approx2.3\sigma$.}
    \label{rel2abs_b201}
  \end{center}
\end{figure}

The median and standard errors on the relative motions of the above selection were $0.45\pm0.17$~mas~yr$^{-1}$ and $1.51\pm0.21$~mas~yr$^{-1}$ in $\alpha\cos\delta$ and $\delta$ respectively. 
The average motions of sources in the Bescan\c{c}on synthetic stellar population model at this Galactic location are $-0.40$~mas~yr$^{-1}$ and $-2.41$~mas~yr$^{-1}$ in $\alpha\cos\delta$ and $\delta$ respectively. 
Since we are comparing Besan\c{c}on average absolute motions to the correction we would need to apply to convert VIRAC relative motions to absolute motions, the two should have equal magnitude and opposite sign. The values agree in $\mu_{\alpha\cos\delta}$ comfortably, but there is some discrepancy between the two in $\mu_{\delta}$. Note that at this Galactic location $\mu_{\alpha\cos\delta}\approx\mu_{b}$ and $\mu_{\delta}\approx\mu_{l}$. Figure \ref{rel2abs_b201} shows the proper motion distributions in $\delta$ in more detail. The spread in the relative motions of the b201 galaxy selection is roughly consistent with their being at the faint end of VIRAC ($K_s>14$).

We obtain the same relative to absolute correction in $\mu_{\alpha\cos\delta}$ for the neighbouring VVV tile, b202 ($l=352.2^{\circ}$; $b=-9.7^{\circ}$). Despite our expectation that it would also remain the same in $\mu_{\delta}$ between close fields, the correction for b202 is $0.71\pm0.18$~mas~yr$^{-1}$. This highlights the difficulty in using suspected external galaxies as relative to absolute calibrators at this level of precision.

\subsection{The Proper Motion of the Sagittarius Dwarf Spheroidal Galaxy}

The Sagittarius Dwarf Spheroidal (Sgr dSph) is a satellite galaxy of the Milky Way identified by \citet{ibata94}. Several measurements of the absolute proper motion of the Sgr dSph exist in the literature, e.g. \citet{dinescu05} ($\mu_{l}\cos{}b=-2.35\pm0.20$ and $\mu_{b}=2.07\pm0.20$~mas~yr$^{-1}$), \citet{pryor10} ($\mu_{l}\cos{}b=-2.61\pm0.22$ and $\mu_{b}=1.87\pm0.19$~mas~yr$^{-1}$), \citet{massari13} ($\mu_{l}\cos{}b=-2.13\pm0.16$ and $\mu_{b}=1.82\pm0.18$~mas~yr$^{-1}$). 

\begin{figure}
  \begin{center}
    \begin{tabular}{c}
      \epsfig{file=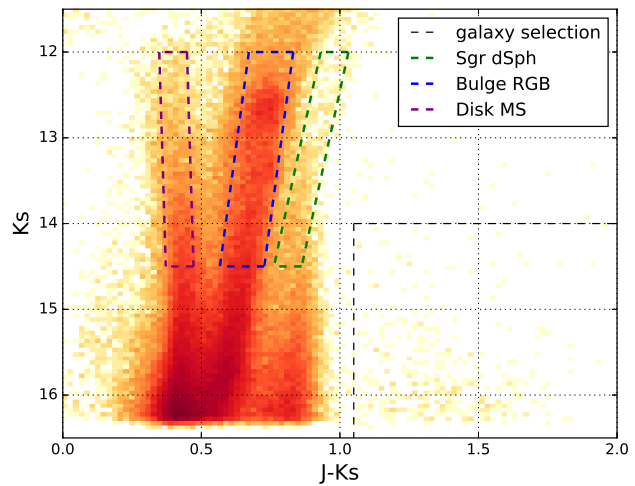,width=1\linewidth,clip=}
    \end{tabular}
    \caption{The $K_s$ vs. $J-K_s$ colour-magnitude diagram for sources flagged as reliable from VVV b211. Highlighted are our selection boxes for external galaxies, the Sgr dSph, Galactic bulge giants and Galactic disk main sequence stars.}
    \label{sgr_dsph_cmd}
  \end{center}
\end{figure}

We attempt a measurement of the absolute proper motion of the Sgr dSph in VVV tile b211 using VIRAC proper motions. We first select external galaxies, Sgr dSph giants, Galactic bulge giants and Galactic disk main sequence stars in $J-K_s$ colour and magnitude (see Figure \ref{sgr_dsph_cmd}).

\begin{figure}
  \begin{center}
    \begin{tabular}{c}
      \epsfig{file=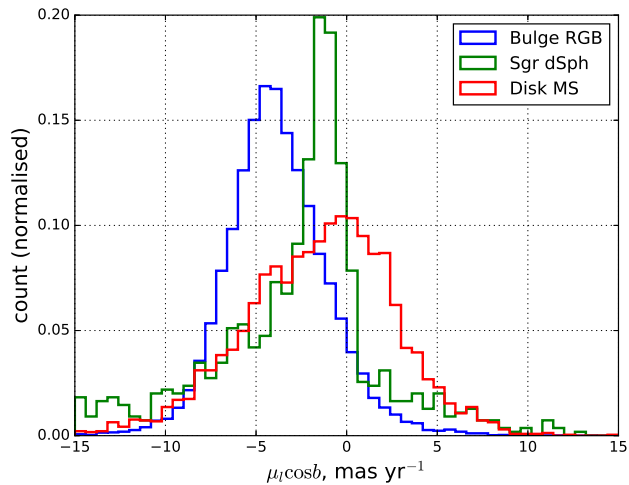,width=1\linewidth,clip=}
    \end{tabular}
    \caption{The absolute proper motion distributions in $l\cos{}b$ of the three stellar selections shown in Figure \ref{sgr_dsph_cmd}.}
    \label{sgr_field_pmdists}
  \end{center}
\end{figure}

\begin{figure}
  \begin{center}
    \begin{tabular}{c}
      \epsfig{file=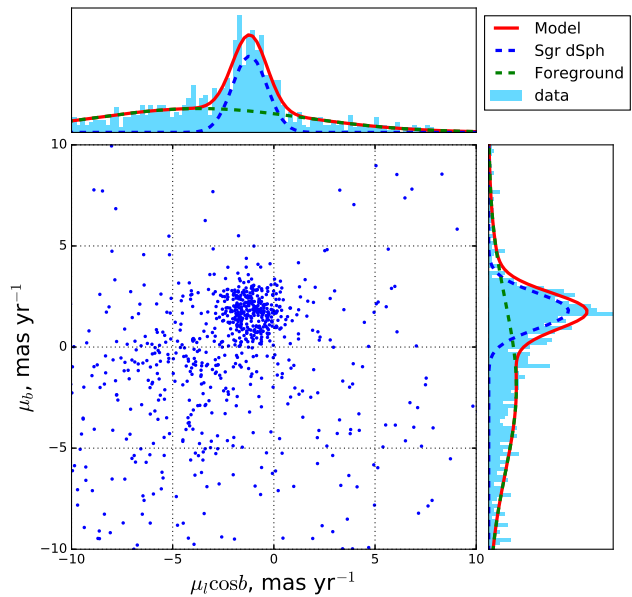,width=1\linewidth,clip=}
    \end{tabular}
    \caption{The proper motion distribution of the Sgr dSph selection. We fit the sum of two Gaussians to the histogram in each dimension to model the sample (red curves). The individual Gaussian fits give us the proper motion and dispersion of the Sgr dSph (blue dashed line), and foreground K and M type dwarfs (green dashed line).}
    \label{sgr_dsph_pm}
  \end{center}
\end{figure}

The external galaxies were used to correct the relative VIRAC proper motions to absolute using the same method as in Section \ref{rel2abs}. We found in Section \ref{rel2abs} that a relative to absolute correction performed in this way should not be trusted implicitly. The corrections applied were $2.92$ and $0.85$~mas~yr$^{-1}$ in $\mu_{l}\cos{}b$ and $\mu_{b}$ respectively, the standard errors on these values are $0.26$ and $0.19$~mas~yr$^{-1}$ respectively but it's quite probable that there are unknown systematic uncertainties as well, we discuss this later.
Figure \ref{sgr_field_pmdists} shows that the three stellar selections have different $\mu_{l}\cos{}b$ distributions in the VIRAC data. In particular, the Sgr dSph selection is clearly peaked, even more so in Figure \ref{sgr_dsph_pm}.

The main panel of Figure \ref{sgr_dsph_pm} shows a locus of sources at $\mu_{l}\cos{b}\approx-1.2$, $\mu_{b}\approx1.8$~mas~yr$^{-1}$ which we interpret to be Sgr dSph sources. The wider spread of sources we interpret as foreground K and M type dwarfs. Histograms of the proper motion distributions in each dimension are shown in the upper and right panels. To each of these distributions we fit the sum of two Gaussians to measure the Sgr dSph proper motion. The Gaussian corresponding to the Sgr dSph sources is describe in $\mu_{l}\cos{}b$ by a mean of $-1.20\pm0.02$~mas~yr$^{-1}$ and a standard deviation of $0.86\pm0.03$~mas~yr$^{-1}$, and in $\mu_{b}$ by a mean of $1.79\pm0.03$~mas~yr$^{-1}$ and a standard deviation of $0.88\pm0.03$~mas~yr$^{-1}$. Ignoring the uncertainty on the relative to absolute correction, the accuracy on Sgr dSph proper motion is encouraging. Once we also take into account the standard error on the relative to absolute correction we find our values have approximately the same accuracy as existing measurements: $\mu_{l}\cos{}b=-1.20\pm0.26$ and $\mu_{b}=1.79\pm0.19$~mas~yr$^{-1}$. Our value for the $\mu_{b}$ agrees with the existing measurements in the literature, the disagreement in $\mu_{l}\cos{}b$ is very likely to be due to an unreliable relative to absolute correction. Another issue is that the existing measurements are of different regions of the Sgr dSph, which will introduce perspective effects due to the imperfect parallelism between the lines of sight \citep{massari13}, but correcting for this is beyond the scope of this paper. 

\subsection{Other uses: Kinematic Distance Estimates and Cluster Decontamination}

In principle, one can use proper motion to estimate distance to an object using a Galactic disc rotation model and assuming the source has a space velocity consistent with disk membership. This is routinely done with radial velocities (e.g. \citealt{contreras17b}). At minimum, proper motions can be used to provide useful constraints on source distances.

\begin{figure*}
  \begin{center}
    \begin{tabular}{c}
      \epsfig{file=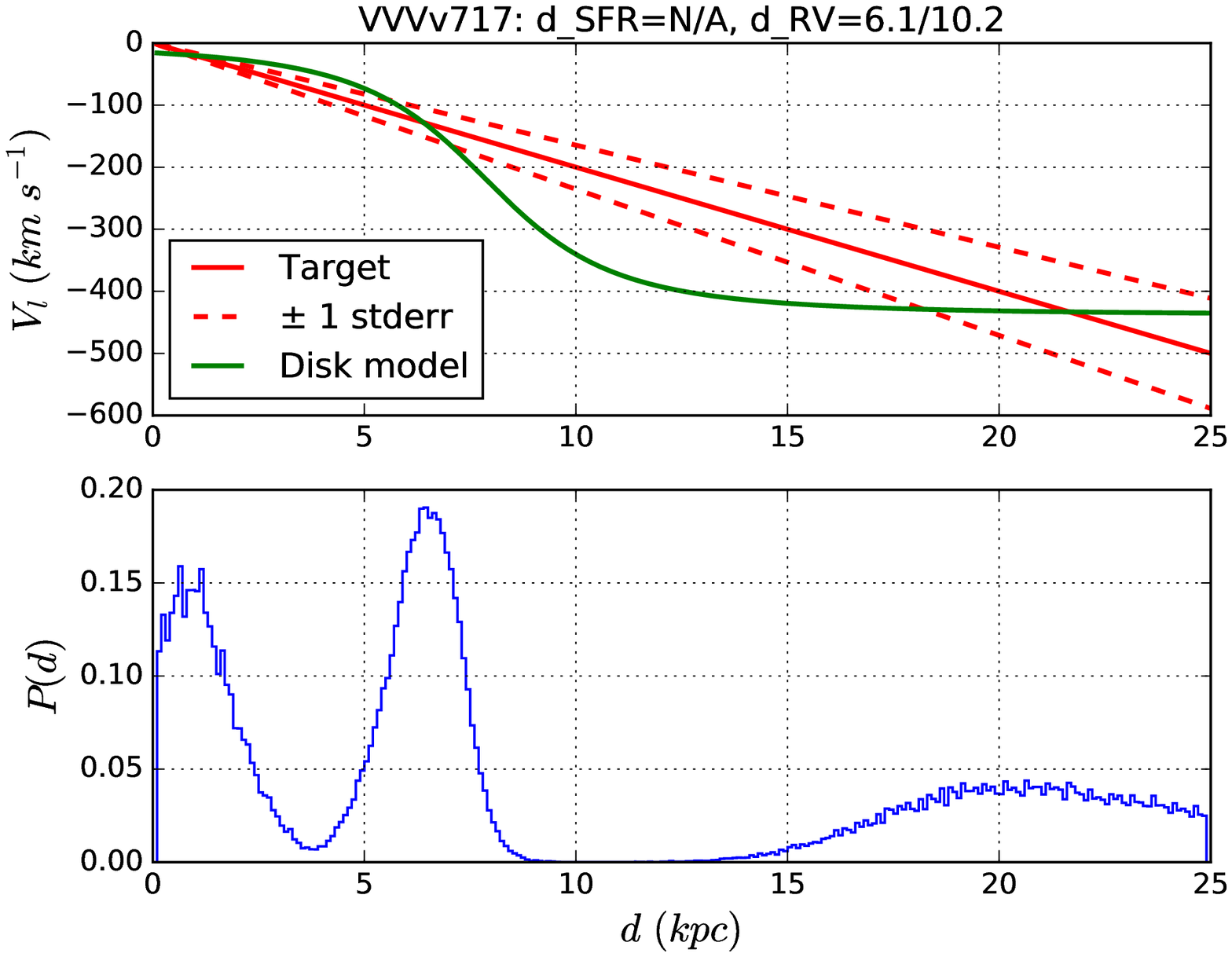,width=0.5\linewidth,clip=}
      \epsfig{file=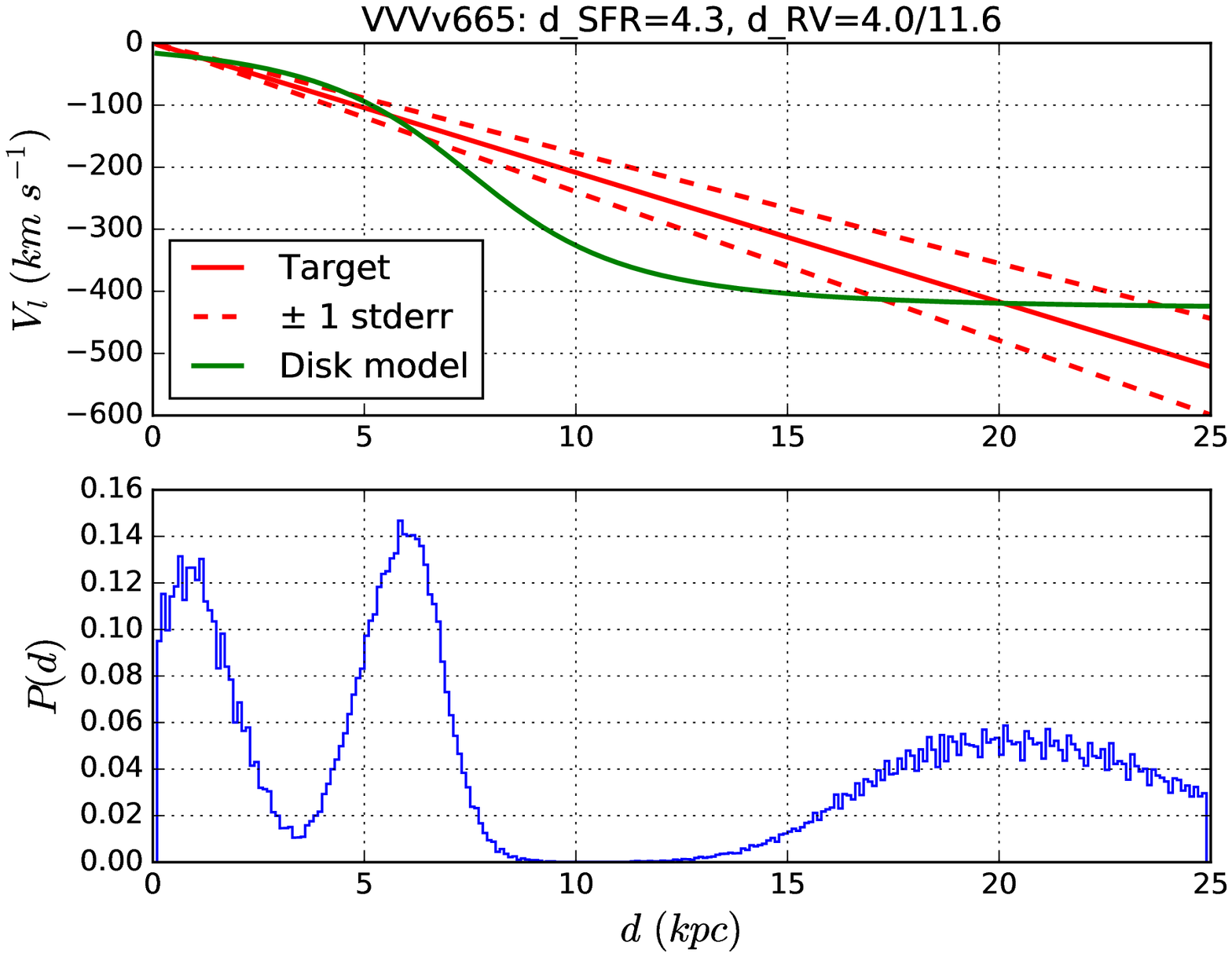,width=0.5\linewidth,clip=}
    \end{tabular}
    \caption{\textit{upper right}: Our Galactic disk rotation model at $l=337.68^{\circ}$, and projected tangential velocity of VVVv717 \citep{contreras17b} using its VIRAC proper motion and uncertainty at the range of heliocentric distances covered by our disk model and a relative to absolute correction based on a Besan\c{c}on synthetic stellar population. \textit{upper left}: The same but for VVVv665 at $l=331.72^{\circ}$. \textit{lower right}: Distance probability distribution of VVVv717 from intersection points between 100,000 monte carlo samples drawn from the VIRAC proper motion and uncertainty, and a peculiar motion of very young (blue) objects of $0\pm15$~km~s$^{-1}$ (\citealt{dehnen98}, both normally distributed). \textit{lower right}: The same but for VVVv665. The \citet{contreras17b} distance estimates for the two sources were based on a similar method but using radial velocity instead of proper motion and also in the case of VVVv665 an estimate based on the distance to a star formation region relatively near to the target (d\_SFR).}
    \label{pm_to_distance_fig}
  \end{center}
\end{figure*}

We give two examples in Figure \ref{pm_to_distance_fig}, for the highly variable YSOs VVVv665 and VVVv717 from \citet{contreras17b}. In both cases we produce a curve of velocity in Galactic longitude versus heliocentric distance using a simple flat Galactic disk rotation model, as in Section \ref{galrotcurve}. We place VIRAC proper motions of the two objects on an absolute frame using a Besan\c{c}on synthetic stellar population and randomly sample from a normal distribution about the absolute proper motion with a standard deviation equal to the proper motion uncertainty. We project this proper motion into a velocity in Galactic longitude and then add a peculiar velocity drawn randomly from a normal distribution with mean $0$~km~s$^{-1}$ and standard deviation $15$~km~s$^{-1}$. The peculiar velocity takes into account velocity dispersion among young objects \citep{dehnen98}. We then note the distances at intersections between the line for projected target velocity and the model, weighting them as the reciprocal of the number of intersections (there may be up to 3 intersections). This is done a total of 100,000 times in a Monte Carlo fashion to produce the probability distributions seen in the lower panels of Figure \ref{pm_to_distance_fig}.

\citet{contreras17b} computed near and far kinematic distances for these two YSOs using the radial velocity of each target and a similar Galactic rotation curve. These are also shown in Figure \ref{pm_to_distance_fig} as "$d_{RV}$" (in kpc), as well as a distance estimate, "$d_{SFR}$" from the literature based on probable association with an adjacent star formation region, in the case of VVVv665.
The proper motion-based distance estimate of VVVv717 agrees well with the near distance from radial velocity. The agreement is less good for VVVv665 but in both cases the far distance from the radial velocity is effectively ruled out using this method.
Again, we stress that these results are highly dependent on a fairly uncertain relative to absolute correction. A more accurate relative to absolute correction, planned for VIRAC version 2 based on Gaia DR2 will make this a more powerful tool. Nonetheless, it appears that applying this method with VIRAC can break the near/far kinematic distance ambiguity given by radial velocities and provide useful constraints on distance in cases where there is no independent estimate.

Another example use of VIRAC for constraining distance is that of \citet{britt16}, where VIRAC proper motion and parallax upper limits were used to determine that the very unusual long-duration transient source CX330 is outside the solar neighbourhood and moving too slowly to be connected to any known star forming complex, assuming an age of order 1~Myr. VIRAC kinematic distance estimates are most useful for objects known to be young and hence very likely to have space velocity consistent with disk rotation, these objects would include young objects (e.g. Cepheids) and open clusters.

A further important use of VIRAC astrometry at large distances is kinematic decontamination of Galactic clusters. A good example of this was the case of the globular cluster FSR~1735 = 2MASS~GC03 at $d = 11$~kpc \citep{carballo-bello16}. They demonstrated that in an early version of the VIRAC proper motions were able to reliably distinguish cluster members on the giant branch from field stars spread more equally across the dwarf and giant branches of the colour magnitude diagram, see figure 5 of that work. See also Section \ref{ngc6231} of this paper.

\section{Summary}

VIRAC V1 comprises near-infrared proper motion and $5\sigma$ parallax catalogues for 560 deg$^{2}$ of the southern Galactic plane and bulge. The proper motion and parallax catalogues contain 312 million sources and 6,935 sources respectively. Sub-mas~yr$^{-1}$ precision on relative proper motions is typical for bright sources, and at the level of a few mas~yr$^{-1}$ at $K_s=16$. Using separate astrometric measurements from overlapping sets of pawprints and also looking at the proper motion dispersion of NGC 6231 cluster members we have demonstrated that these uncertainties characterise the true statistical errors well.
We present 18 new L dwarf candidates with parallax measurements and a further 66 identified through their high proper motion. We have found a valuable chemical abundance benchmark L dwarf and a very rare L subdwarf with very blue colours and estimated metallicity [Fe/H]$\sim -1$. The parallax catalogue includes ten sources with $d<25$~pc, nine of which are new discoveries.

We have demonstrated that VIRAC astrometry is useful for objects at a large Galactic distance through: a measurement of the Galactic rotation curve at $l=300^{\circ}$ out to $d \approx 12$~kpc; a measurement of the absolute proper motion of the Sagittarius dwarf spheroidal galaxy in VVV tile b211; a measurement of the absolute proper motion of the Galactic bulge in VVV tile b201; and kinematic distance measurements of two young stellar objects. Due to some doubt over the accuracy of the relative to absolute proper motion correction in these fields these results at large Galactic distances are very much preliminary. A precise relative to absolute correction based on the forthcoming Gaia 2nd data release and VIRAC V2 will enable more serious work on these topics.

Due to the high extinction in optical bandpasses in the Galactic mid-plane and bulge, VIRAC will continue to retain value in the era of Gaia and beyond.

\section*{Acknowledgments}
We are grateful to Dr Floor van Leeuwen for a very helpful referee report, and to Dr Katelyn Allers and Sean Points (ARCoIRIS instrument scientist) for assistance during spectroscopic observation of LTT 7251 B. LCS acknowledges a studentship funded by the Science \& Technology Facilities Research Council (STFC) of the UK; LCS, PWL, HRAJ, FM, JED and DJP acknowledge the support of a consolidated grant (ST/J001333/1 and ST/M001008/1) also funded by STFC. We acknowledge use of data from the ESO Public Survey programme ID 179.B-2002 taken with the VISTA telescope, data products from CASU, and funding from the FONDAP Center for Astrophysics 15010003, the BASAL CATA Center for Astrophysics and Associated Technologies PFB-06, the FONDECYT from CONICYT.

\bibliographystyle{mn2e}
\bibliography{main}

\onecolumn
\appendix
\begin{landscape}
\section{Confirmed High Proper Motion sources}
\begin{table}
\caption{High proper motion sources confirmed by visual inspection. \textit{Full table in online data.}}
\label{hpmtable}
\begin{tabular}{|r|l|l|l|r|r|r|r|r|r|}
\hline
  \multicolumn{1}{|c|}{\#} &
  \multicolumn{1}{c|}{$\alpha_{2012.0}$} &
  \multicolumn{1}{c|}{$\delta_{2012.0}$} &
  \multicolumn{1}{c|}{$\mu_{\alpha\cos\delta}$} &
  \multicolumn{1}{c|}{$\mu_{\delta}$} &
  \multicolumn{1}{c|}{Z} &
  \multicolumn{1}{c|}{Y} &
  \multicolumn{1}{c|}{J} &
  \multicolumn{1}{c|}{H} &
  \multicolumn{1}{c|}{Ks} \\
\hline
  1 & 11h38m53.05s  &  -63d46m04.54s  &  $ -486.45\pm 0.88$  &  $    9.72\pm 0.77$  &  $13.27\pm0.01$  &  $12.71\pm0.01$  &  $11.71\pm0.01$  &  $11.39\pm0.01$  &  $11.35\pm0.01$ \\ 
  2 & 11h39m17.78s  &  -63d00m41.58s  &  $ -211.71\pm 0.83$  &  $  122.61\pm 0.74$  &  $14.62\pm0.01$  &  $14.01\pm0.01$  &  $13.38\pm0.01$  &  $13.06\pm0.01$  &  $12.69\pm0.02$ \\ 
  3 & 11h41m44.65s  &  -63d28m26.79s  &  $ -224.34\pm 0.88$  &  $  109.19\pm 0.78$  &  $12.91\pm0.01$  &  $12.57\pm0.01$  &  $12.40\pm0.01$  &  $12.56\pm0.01$  &  $11.30\pm0.07$ \\ 
  4 & 11h46m27.36s  &  -60d36m40.41s  &  $  -23.34\pm 2.19$  &  $ -219.78\pm 2.23$  &  $11.75\pm0.01$  &  $11.26\pm0.01$  &  $10.62\pm0.01$  &  $10.30\pm0.01$  &  $10.10\pm0.07$ \\ 
  5 & 11h51m51.42s  &  -59d57m29.86s  &  $  -23.97\pm 0.89$  &  $  266.86\pm 0.87$  &  $14.49\pm0.01$  &  $13.80\pm0.01$  &  $13.18\pm0.01$  &  $12.42\pm0.01$  &  $12.46\pm0.01$ \\ 
  6 & 11h53m04.81s  &  -62d01m44.55s  &  $ -262.73\pm 0.75$  &  $   10.02\pm 0.81$  &  $13.70\pm0.01$  &  $13.06\pm0.01$  &  $12.39\pm0.01$  &  $11.48\pm0.01$  &  $11.59\pm0.01$ \\ 
  7 & 11h54m03.86s  &  -60d05m29.35s  &  $ -565.55\pm 0.85$  &  $  216.76\pm 0.83$  &  $13.63\pm0.01$  &  $12.99\pm0.01$  &  $12.37\pm0.01$  &  $11.62\pm0.01$  &  $11.64\pm0.01$ \\
\hline\end{tabular}
\vspace{10cm}
\end{table}
\clearpage

\section{Parallax Sources}
\begin{table}
\caption{Parallax sources measured at $5\sigma$ \textit{Full table in online data.}}
\label{plxtable}
\begin{tabular}{|r|l|l|l|r|r|r|r|r|r|r|}
\hline
  \multicolumn{1}{|c|}{\#} &
  \multicolumn{1}{c|}{$\alpha_{2012.0}$} &
  \multicolumn{1}{c|}{$\delta_{2012.0}$} &
  \multicolumn{1}{c|}{$\mu_{\alpha\cos\delta}$} &
  \multicolumn{1}{c|}{$\mu_{\delta}$} &
  \multicolumn{1}{c|}{$\pi$} &
  \multicolumn{1}{c|}{Z} &
  \multicolumn{1}{c|}{Y} &
  \multicolumn{1}{c|}{J} &
  \multicolumn{1}{c|}{H} &
  \multicolumn{1}{c|}{Ks} \\
\hline
   1 & 11h38m53.06s  &  -63d46m04.54s  &  $ -482.00\pm 0.70$  &  $   13.14\pm 0.71$  &  $16.9\pm1.7$  &  $13.27\pm0.01$  &  $12.71\pm0.01$  &  $11.71\pm0.01$  &  $11.39\pm0.01$  &  $11.35\pm0.01$ \\ 
   2 & 11h39m15.84s  &  -63d03m36.99s  &  $ -115.79\pm 0.77$  &  $   30.45\pm 0.72$  &  $15.2\pm1.7$  &  $15.62\pm0.01$  &  $14.67\pm0.01$  &  $13.77\pm0.01$  &  $13.20\pm0.01$  &  $12.75\pm0.01$ \\ 
   3 & 11h39m17.78s  &  -63d00m41.58s  &  $ -210.52\pm 0.65$  &  $  121.91\pm 0.72$  &  $14.6\pm1.6$  &  $14.62\pm0.01$  &  $14.01\pm0.01$  &  $13.38\pm0.01$  &  $13.06\pm0.01$  &  $12.69\pm0.02$ \\ 
   4 & 11h39m21.52s  &  -63d29m02.18s  &  $ -140.25\pm 0.49$  &  $   16.96\pm 0.58$  &  $16.3\pm1.3$  &  $13.54\pm0.01$  &  $13.04\pm0.01$  &  $12.10\pm0.01$  &  $11.67\pm0.01$  &  $11.68\pm0.01$ \\ 
   5 & 11h40m08.73s  &  -62d32m43.31s  &  $ -158.83\pm 0.68$  &  $   45.83\pm 0.89$  &  $11.3\pm1.8$  &  $13.24\pm0.01$  &  $12.73\pm0.01$  &  $11.72\pm0.01$  &  $11.34\pm0.01$  &  $11.32\pm0.01$ \\ 
   6 & 11h40m34.75s  &  -63d14m04.92s  &  $  -71.43\pm 0.60$  &  $   11.48\pm 0.60$  &  $ 8.2\pm1.4$  &  $13.97\pm0.01$  &  $13.43\pm0.01$  &  $12.81\pm0.01$  &  non detection  &  $11.97\pm0.01$ \\ 
   7 & 11h40m43.80s  &  -63d14m41.62s  &  $   25.90\pm 0.75$  &  $   43.91\pm 0.77$  &  $10.2\pm1.8$  &  $16.36\pm0.01$  &  $15.63\pm0.01$  &  $14.86\pm0.01$  &  $14.34\pm0.01$  &  $13.96\pm0.01$ \\
\hline\end{tabular}
\vspace{10cm}
\end{table}

\clearpage

\section{Solar Neighbourhood Sources}
\begin{table}
\caption{VIRAC sources whose parallaxes place them within 25~pc.}
\label{dlt25table}
\begin{tabular}{|r|l|l|l|r|r|r|r|r|r|r|}
\hline
  \multicolumn{1}{|c|}{\#} &
  \multicolumn{1}{c|}{$\alpha_{2012.0}$} &
  \multicolumn{1}{c|}{$\delta_{2012.0}$} &
  \multicolumn{1}{c|}{$\mu_{\alpha\cos\delta}$} &
  \multicolumn{1}{c|}{$\mu_{\delta}$} &
  \multicolumn{1}{c|}{$\pi$} &
  \multicolumn{1}{c|}{Z} &
  \multicolumn{1}{c|}{Y} &
  \multicolumn{1}{c|}{J} &
  \multicolumn{1}{c|}{H} &
  \multicolumn{1}{c|}{Ks} \\
\hline
   1 & 13h34m27.67s  &  -62d57m35.90s  &  $ -213.12\pm 1.02$  &  $  -21.98\pm 0.86$  &  $52.7\pm2.1$  &  $14.81\pm0.01$  &  $13.60\pm0.01$  &  $12.51\pm0.01$  &  $11.56\pm0.01$  &  $11.37\pm0.01$ \\ 
   2 & 14h11m59.32s  &  -59d20m45.70s  &  $   82.15\pm 0.80$  &  $  -54.60\pm 0.98$  &  $66.8\pm1.7$  &  $13.91\pm0.01$  &  $13.96\pm0.01$  &  $13.79\pm0.01$  &  $13.64\pm0.01$  &  $13.55\pm0.01$ \\ 
   3 & 17h06m28.37s  &  -32d37m33.41s  &  $ -151.35\pm 2.04$  &  $ -100.22\pm 2.14$  &  $45.1\pm5.1$  &  $12.31\pm0.01$  &  $11.36\pm0.01$  &  $10.76\pm0.01$  &  $10.34\pm0.01$  &  $ 9.91\pm0.03$ \\ 
   4 & 17h13m40.60s  &  -39d52m15.21s  &  $  111.89\pm 0.61$  &  $ -281.55\pm 0.62$  &  $48.7\pm1.6$  &  $15.59\pm0.01$  &  $14.45\pm0.01$  &  $13.30\pm0.01$  &  $12.20\pm0.01$  &  $12.09\pm0.01$ \\ 
   5 & 17h26m40.15s  &  -27d38m03.72s  &  $ -544.39\pm 0.48$  &  $ -323.48\pm 0.65$  &  $53.3\pm1.1$  &  $15.43\pm0.01$  &  $14.29\pm0.01$  &  $13.20\pm0.01$  &  $12.67\pm0.01$  &  $12.21\pm0.01$ \\ 
   6 & 17h36m41.69s  &  -34d25m47.93s  &  $ -121.43\pm 1.12$  &  $ -139.62\pm 1.19$  &  $54.0\pm1.9$  &  $13.03\pm0.01$  &  $12.27\pm0.01$  &  $10.99\pm0.01$  &  $10.46\pm0.01$  &  $10.09\pm0.04$ \\ 
   7 & 17h43m49.16s  &  -39d08m25.32s  &  $ -113.59\pm 0.45$  &  $   52.32\pm 0.48$  &  $44.9\pm1.2$  &  $13.52\pm0.01$  &  $13.51\pm0.01$  &  $13.62\pm0.01$  &  $13.61\pm0.01$  &  $13.49\pm0.01$ \\
\hline\end{tabular}
\vspace{10cm}
\end{table}

\end{landscape}
\clearpage

\section{VIRAC - TGAS Common Proper Motion sources}
\begin{table}
\vspace{-10cm}
\tiny
\caption{VIRAC common proper motion companions to TGAS high proper motion sources.}
\label{tgas_companions}
\begin{tabular}{|l|l|l|r|r|r|r|r|r|r|r|}
\hline
  \multicolumn{1}{|c|}{Name} &
  \multicolumn{1}{c|}{$\alpha_{2015.0}$} &
  \multicolumn{1}{c|}{$\delta_{2015.0}$} &
  \multicolumn{1}{c|}{$\mu_{\alpha\cos\delta}$} &
  \multicolumn{1}{c|}{$\mu_{\delta}$} &
  \multicolumn{1}{c|}{$\pi$} &
  \multicolumn{1}{c|}{Gaia G Mag} &
  \multicolumn{1}{c|}{Ks} &
  \multicolumn{1}{c|}{MKs} &
  \multicolumn{1}{c|}{$\rho$ \arcsec} &
  \multicolumn{1}{c|}{$\rho (au)$} \\
\hline
  HD 147436 A & 16h24m13.46s & -47d35m16.3s & $-79.02\pm0.11$ & $-72.57\pm0.07$ & $16.71\pm0.24$ & 9.04 & $7.51\pm0.02$ & 3.62 &  & \\
  HD 147436 B & 16h24m13.24s & -47d35m15.4s & $-75.72\pm2.83$ & $-74.59\pm2.82$ &  &  & $11.12\pm0.02$ & 7.24 & 2.4 & 143 \\
  &&&&&&&&&&\\
  TYC 8990-1546-1 A & 13h17m50.13s & -60d41m11.1s & $151.11\pm0.59$ & $-51.80\pm0.77$ & $22.44\pm0.26$ & 9.74 & $7.51\pm0.03$ & 4.26 &  & \\
  TYC 8990-1546-1 B & 13h17m50.44s & -60d41m13.0s & $156.11\pm1.64$ & $-47.35\pm1.81$ &  &  & $10.02\pm0.10$ & 6.78 & 3.0 & 135 \\
  &&&&&&&&&&\\
  TYC 8975-2832-1 A & 12h24m39.97s & -61d23m21.5s & $-116.11\pm0.72$ & $-27.73\pm1.14$ & $6.18\pm0.33$ & 9.72 & $7.98\pm0.02$ & 1.93 &  & \\
\hline\end{tabular}
\vspace{3cm}
\end{table}
\clearpage
\begin{table}
\tiny
\caption{TGAS - VIRAC common proper motion and parallax companions.}
\label{tgas_cpm_plx}
\begin{tabular}{|l|l|l|r|r|r|r|r|r|r|r|}
\hline
  \multicolumn{1}{|c|}{Name} &
  \multicolumn{1}{c|}{$\alpha_{2015.0}$} &
  \multicolumn{1}{c|}{$\delta_{2015.0}$} &
  \multicolumn{1}{c|}{$\mu_{\alpha\cos\delta}$} &
  \multicolumn{1}{c|}{$\mu_{\delta}$} &
  \multicolumn{1}{c|}{$\pi$} &
  \multicolumn{1}{c|}{Gaia G Mag} &
  \multicolumn{1}{c|}{Ks} &
  \multicolumn{1}{c|}{MKs} &
  \multicolumn{1}{c|}{$\rho$ \arcsec} &
  \multicolumn{1}{c|}{$\rho (au)$} \\
\hline
Tycho2 8974-1133-1 A & 12h08m08.33s  &  -60d46m51.90s  &  $  -87.13\pm 0.29$  &  $   31.21\pm 0.29$  &  $22.3\pm0.6$  &  $ 7.33$  &     &     &     &    \\
Tycho2 8974-1133-1 B & 12h08m10.82s  &  -60d46m43.60s  &  $  -80.08\pm 0.61$  &  $   29.17\pm 0.61$  &  $18.1\pm1.8$  &    & $11.80\pm0.01$ &  $  8.54\pm  0.06$ &   $ 20.0$    &  $   896$  \\ 
 &  &  &  &  &  &  &  &  &  & \\
Tycho2 8982-1530-1 A & 12h10m51.52s  &  -64d25m28.34s  &  $ -116.45\pm 0.84$  &  $  -45.93\pm 0.58$  &  $15.7\pm0.3$  &  $ 9.41$  &     &     &     &    \\
Tycho2 8982-1530-1 B & 12h10m50.03s  &  -64d24m43.68s  &  $ -108.36\pm 0.59$  &  $  -46.67\pm 0.71$  &  $17.6\pm1.6$  &    & $12.20\pm0.01$ &  $  8.18\pm  0.05$ &   $ 45.7$    &  $  2911$  \\ 
 &  &  &  &  &  &  &  &  &  & \\
Tycho2 8975-2626-1 A & 12h27m32.49s  &  -61d14m27.50s  &  $  -62.35\pm 1.36$  &  $    8.12\pm 0.65$  &  $ 8.7\pm0.3$  &  $ 9.97$  &     &     &     &    \\
\hline\end{tabular}
    \end{table}

    \begin{table}
    \tiny
    \contcaption{}
    \begin{tabular}{|l|l|l|r|r|r|r|r|r|r|r|}
    \hline
      \multicolumn{1}{|c|}{Name} &
      \multicolumn{1}{c|}{$\alpha_{2015.0}$} &
      \multicolumn{1}{c|}{$\delta_{2015.0}$} &
      \multicolumn{1}{c|}{$\mu_{\alpha\cos\delta}$} &
      \multicolumn{1}{c|}{$\mu_{\delta}$} &
      \multicolumn{1}{c|}{$\pi$} &
      \multicolumn{1}{c|}{Gaia G Mag} &
      \multicolumn{1}{c|}{Ks} &
      \multicolumn{1}{c|}{MKs} &
      \multicolumn{1}{c|}{$\rho$ \arcsec} &
      \multicolumn{1}{c|}{$\rho (au)$} \\
    \hline
    Hip 88440 A & 18h03m21.99s  &  -25d54m60.00s  &  $   -8.15\pm 0.09$  &  $  -62.72\pm 0.06$  &  $10.6\pm0.2$  &  $ 8.18$  &     &     &     &    \\
    Hip 88440 B & 18h03m23.44s  &  -25d55m14.15s  &  $   -7.53\pm 0.40$  &  $  -56.15\pm 0.34$  &  $ 9.5\pm1.0$  &    & $12.44\pm0.01$ &  $  7.57\pm  0.05$ &   $ 24.2$    &  $  2281$  \\ 
 &  &  &  &  &  &  &  &  &  & \\
Tycho2 6259-1989-1 A & 18h03m26.00s  &  -19d13m37.74s  &  $   29.83\pm 2.02$  &  $    5.09\pm 1.58$  &  $ 5.4\pm0.8$  &  $10.81$  &     &     &     &    \\
Tycho2 6259-1989-1 B & 18h03m27.37s  &  -19d13m20.14s  &  $   34.12\pm 0.78$  &  $    7.79\pm 0.70$  &  $ 8.2\pm1.2$  &    & $13.95\pm0.01$ &  $  7.62\pm  0.32$ &   $ 26.1$    &  $  4833$  \\ 
 &  &  &  &  &  &  &  &  &  & \\
Tycho2 7391-2148-1 A & 18h04m17.60s  &  -31d33m25.28s  &  $    0.06\pm 2.14$  &  $  -45.13\pm 2.19$  &  $ 5.5\pm0.2$  &  $10.07$  &     &     &     &    \\
\hline\end{tabular}
\end{table}

\clearpage

\begin{landscape}
\section{L and Early T Dwarf Candidates}
\begin{table}
\tiny
\caption{L0-T2 dwarf candidates from our parallax catalogue based on NIR colours and Ks band luminosity.}
\label{ldwftable}
\begin{tabular}{|r|l|l|l|r|r|r|r|r|r|r|r|}
\hline
  \multicolumn{1}{|c|}{\#} &
  \multicolumn{1}{c|}{$\alpha_{2012.0}$} &
  \multicolumn{1}{c|}{$\delta_{2012.0}$} &
  \multicolumn{1}{c|}{$\mu_{\alpha\cos\delta}$} &
  \multicolumn{1}{c|}{$\mu_{\delta}$} &
  \multicolumn{1}{c|}{$\pi$} &
  \multicolumn{1}{c|}{Z} &
  \multicolumn{1}{c|}{Y} &
  \multicolumn{1}{c|}{J} &
  \multicolumn{1}{c|}{H} &
  \multicolumn{1}{c|}{Ks} &
  \multicolumn{1}{c|}{$M_{Ks}$} \\
\hline
 1 & 12h27m08.17s  &  -62d14m20.07s  &  $  -29.61\pm 0.83$  &  $  -11.73\pm 0.71$  &  $ 14.03\pm 2.19$  &  $17.92\pm0.02$  &  $16.75\pm0.01$  &  $15.76\pm0.01$  &  $15.13\pm0.01$  &  $14.53\pm0.01$  &  $10.27\pm0.34$ \\ 
 2 & 13h26m57.62s  &  -61d18m18.13s  &  $ -199.38\pm 0.62$  &  $  -64.37\pm 0.71$  &  $ 15.17\pm 1.53$  &  $17.03\pm0.01$  &  $15.97\pm0.01$  &  $15.07\pm0.01$  &  $14.44\pm0.01$  &  $13.96\pm0.01$  &  $ 9.87\pm0.22$ \\ 
 3 & 13h36m47.45s  &  -61d46m48.48s  &  $  129.02\pm 0.85$  &  $   55.86\pm 0.93$  &  $ 30.95\pm 1.89$  &  $18.49\pm0.03$  &  $17.21\pm0.02$  &  $15.93\pm0.01$  &  $15.13\pm0.01$  &  $14.55\pm0.01$  &  $12.00\pm0.13$ \\ 
 4 & 14h01m16.88s  &  -60d55m53.74s  &  $ -121.15\pm 1.49$  &  $ -131.96\pm 1.62$  &  $ 30.90\pm 3.37$  &  $20.49\pm0.12$  &  $18.90\pm0.06$  &  $17.20\pm0.03$  &  $16.04\pm0.02$  &  $15.05\pm0.02$  &  $12.50\pm0.24$ \\ 
 5 & 14h07m05.64s  &  -62d46m55.76s  &  $ -144.46\pm 1.85$  &  $ -121.93\pm 1.60$  &  $ 30.39\pm 3.73$  &  $19.43\pm0.07$  &  $18.16\pm0.04$  &  $16.84\pm0.02$  &  $15.97\pm0.02$  &  $15.38\pm0.03$  &  $12.80\pm0.27$ \\ 
 6 & 15h03m59.87s  &  -60d28m14.37s  &  $   22.77\pm 0.95$  &  $   -5.41\pm 1.20$  &  $ 13.78\pm 2.08$  &  $18.02\pm0.03$  &  $16.92\pm0.02$  &  $15.82\pm0.01$  &  $15.15\pm0.01$  &  $14.60\pm0.01$  &  $10.30\pm0.33$ \\
\hline\end{tabular}
\vspace{10cm}
\end{table}
\end{landscape}
\begin{table}
\tiny
\caption{Early L dwarf candidates from our proper motion catalogue selected based on $\mu>30~mas~yr^{-1}$ and NIR colours.}
\label{Ltable2}
\begin{tabular}{|r|l|l|l|r|r|r|r|r|r|}
\hline
  \multicolumn{1}{|c|}{\#} &
  \multicolumn{1}{c|}{$\alpha_{2012.0}$} &
  \multicolumn{1}{c|}{$\delta_{2012.0}$} &
  \multicolumn{1}{c|}{$\mu_{\alpha\cos\delta}$} &
  \multicolumn{1}{c|}{$\mu_{\delta}$} &
  \multicolumn{1}{c|}{Z} &
  \multicolumn{1}{c|}{Y} &
  \multicolumn{1}{c|}{J} &
  \multicolumn{1}{c|}{H} &
  \multicolumn{1}{c|}{Ks} \\
\hline
 1 & 12h07m52.19s  &  -63d07m31.76s  &  $  -14.21\pm 1.69$  &  $  -54.81\pm 1.72$  &  $18.14\pm0.03$  &  $17.11\pm0.02$  &  $16.18\pm0.01$  &  $15.58\pm0.01$  &  $15.11\pm0.02$ \\ 
 2 & 12h08m19.98s  &  -60d38m00.38s  &  $   49.26\pm 1.50$  &  $  -33.90\pm 2.14$  &  $18.39\pm0.03$  &  $17.60\pm0.03$  &  $16.56\pm0.01$  &  $15.85\pm0.01$  &  $15.35\pm0.03$ \\ 
 3 & 12h16m44.53s  &  -60d47m45.95s  &  $  -31.26\pm 0.86$  &  $   -9.20\pm 0.78$  &  $17.61\pm0.02$  &  $16.26\pm0.01$  &  $15.18\pm0.01$  &  $14.54\pm0.01$  &  $13.94\pm0.01$ \\ 
 4 & 12h21m09.82s  &  -62d14m40.77s  &  $  -32.95\pm 1.43$  &  $  -16.40\pm 1.88$  &  $18.85\pm0.04$  &  $17.44\pm0.02$  &  $16.23\pm0.01$  &  $15.61\pm0.01$  &  $15.08\pm0.02$ \\ 
 5 & 12h27m08.16s  &  -62d14m20.07s  &  $  -30.36\pm 0.93$  &  $  -11.78\pm 0.74$  &  $17.92\pm0.02$  &  $16.75\pm0.01$  &  $15.76\pm0.01$  &  $15.13\pm0.01$  &  $14.53\pm0.01$ \\ 
 6 & 12h31m31.06s  &  -62d26m31.71s  &  $  -32.81\pm 1.10$  &  $  -12.97\pm 1.02$  &  $18.21\pm0.03$  &  $16.77\pm0.01$  &  $15.63\pm0.01$  &  $15.02\pm0.01$  &  $14.44\pm0.01$ \\ 
 7 & 12h50m03.57s  &  -61d53m51.73s  &  $  -54.69\pm 1.88$  &  $  -18.14\pm 1.51$  &  $19.07\pm0.04$  &  $17.83\pm0.03$  &  $16.67\pm0.01$  &  $15.93\pm0.02$  &  $15.35\pm0.02$ \\
\hline\end{tabular}
\end{table}

\end{document}